\documentclass[onecolumn,floatfix,superscriptaddress,a4paper,
               showpacs,showkeys,nofootinbib,preprint]{revtex4}

\usepackage{epsfig}
\usepackage{latexsym}
\usepackage{xspace}
\usepackage{hyperref}
\usepackage[latin2]{inputenc}
\usepackage{indentfirst}
\usepackage{enumerate}
\usepackage[usenames]{color}

\usepackage{amsmath}
\usepackage{amssymb}
\usepackage[english]{babel}

\usepackage{url}

\newcommand{\bs}{\boldsymbol}

\newcommand*{\dis}{\displaystyle}

\begin{document}

\title{Ideal Boson Particle-Antiparticle System at Finite Temperatures}
\author{D. Anchishkin}
\affiliation{Bogolyubov Institute for Theoretical Physics, 03143 Kyiv, Ukraine}
\affiliation{Taras Shevchenko National University of Kyiv, 03022 Kyiv, Ukraine}
\affiliation{Frankfurt Institute for Advanced Studies,
60438 Frankfurt am Main, Germany}
\author{V. Gnatovskyy}
\affiliation{Bogolyubov Institute for Theoretical Physics, 03143 Kyiv, Ukraine}
\affiliation{Taras Shevchenko National University of Kyiv, 03022 Kyiv, Ukraine}
\author{I. Kondakova}
\affiliation{Frantsevich Institute for Problems of Materials Science,
03142 Kyiv, Ukraine}

\date{\today}



\begin{abstract}
The thermodynamic properties of an ideal bosonic system composed of particles
and antiparticles at finite temperatures are examined within the framework of
a scalar field model.
It is assumed that particle-antiparticle pair creation occurs; however, the
system is simultaneously subject to exact charge (isospin) conservation.

To implement this constraint, we first consider the system within the
Grand Canonical Ensemble and then transform to the Canonical Ensemble using
a Legendre transformation.
This procedure provides a formally consistent scheme for incorporating the
chemical potential at the microscopic level into the Canonical Ensemble
framework.
To enforce exact conservation of charge (isospin, $N_I$), we further analyze
the thermodynamic properties of the system within the extended Canonical
Ensemble, in which the chemical potential becomes a thermodynamic function of
the temperature and conserved charge.
It is shown that as the temperature decreases, the system undergoes a
second-order phase transition to a Bose-Einstein condensate at the critical
temperature $T_{\rm c}$, but only when the conserved charge is finite,
$N_I =$~const~$\ne 0$.
In a particle-antiparticle system, the condensate forms exclusively in the
component with the dominant particle number density, which determines the
excess charge.
We demonstrate that the symmetry breaking of the ground state at $T = 0$
results from a first-order phase transition associated with the formation
of a Bose-Einstein condensate.
Although the transition involves symmetry breaking, it is not spontaneous in
the strict field-theoretic sense, but is instead induced by the external
injection of particles.
Potential experimental signals of Bose-Einstein condensation of pions produced
in high-energy nuclear collisions are briefly discussed.
\end{abstract}

\maketitle

\section{Introduction}
\label{sec1}

In recent years, the properties of hot and dense hadronic matter have
attracted significant interest.
Such matter can be produced in relativistic nucleus-nucleus collisions, which
many laboratories are currently investigating.
Effective models based on QCD and lattice simulations \cite{brandt-2016}
suggest that chiral symmetry restoration and the deconfinement phase
transition should occur at high temperatures and particle densities.
The study of hot and dense meson systems is essential for understanding
hadronic matter in extreme conditions ~\cite{bzdak-esum-2020}.
For example, the multi-pion states created in high-energy collisions of
nuclei and particles can provide signals about the fireball's intrinsic
states and its evolution.
Investigating meson (boson) systems possesses specific characteristics due
to the possibility of Bose-Einstein condensation of bosonic particles.
The connection between spontaneous symmetry breaking in the boson system and
the emergence of the Bose-Einstein condensate is the subject of numerous
theoretical studies.
One of the pioneering works in this area is Ref.~\cite{kapusta-1981}, which
examines phase transitions and the characteristics of the Bose-Einstein
condensate (BEC) in models of non-interacting and self-interacting charged
scalar fields.
The conditions for the emergence of BEC are closely related to the behavior
of the boson chemical potential, regarded as an effective parameter of symmetry
breaking.
In the present paper, within the framework of the scalar field model, we
consider an ideal Bose gas of relativistic particles and antiparticles,
conserving isospin (charge).\footnote{It is important to note that in the
$\pi^- - \pi^+$ system, which we use as
a reference, the isospin states coincide with the charge states.}
The systems of relativistic bosons and Bose-Einstein condensation have been
explored in several papers; for instance, see \cite{beckmann-1979} and the
references therein.
For the first time, charge conservation in an ideal relativistic boson system
of particles and antiparticles was studied in Ref.~\cite{haber-1981}.
The authors derived expansions for the main thermodynamic quantities at
temperatures exceeding the critical temperature $T_{\rm c}$ of the
Bose-Einstein condensation.
It should be noted that describing systems of Bose particles and antiparticles
is typically conducted using a Grand Canonical Ensemble in which the chemical
potential $\mu_I$ serves as a canonical variable.
However, as we will discuss subsequently, in the presence of BEC, the chemical
potential cannot be treated as a free variable; the necessary condition for
condensate formation dictates its value.
The fact that the chemical potential is a dependent parameter that must be
calculated from the known charge density was emphasized in Ref.~\cite{haber-1982}.
The total number of particles in the relativistic system is not conserved
since at high temperatures, particle-antiparticle pairs are created.
Meanwhile, the difference between the number of particles and antiparticles,
$N_I$, or the system's charge, remains constant.
In the Canonical Ensemble, the charge value can be accurately fixed and
considered when calculating thermodynamic quantities.
However, to maintain the quantity $N_I$ constant over a wide temperature range,
the chemical potential $\mu_I$ must be introduced, which requires us to operate
within the framework of the Grand Canonical Ensemble.
To resolve this apparent contradiction, a standard field-theoretic method
was proposed:
introducing the operator of the conserved charge multiplied by the chemical
potential into the Lagrangian and evaluating the statistics from a functional
integral approach \cite{kapusta-1981,haber-1982}.
In 2002, it was demonstrated in Ref.~\cite{salasnich-2002} that this
method first involves applying the effective Hamiltonian within
the Grand Canonical Ensemble, and then moving to the effective Lagrangian
used in statistical description through the Legendre transformation.
However, Ref.~\cite{haber-1982} noted that the chemical potential is a
dependent parameter that must be determined from the known charge density.
In our work, we adopt a similar approach to that in
Ref.~\cite{salasnich-2002}: we begin by formulating the problem within the
Grand Canonical Ensemble and then, using the Legendre transformation,
shift the study to the Canonical Ensemble.
This process results in a thermodynamic description of the boson system
within the Canonical Ensemble, utilizing the chemical potential as a
thermodynamic function that depends on the conserved charge $N_I$ and
temperature $T$.

Various problems related to the ideal and interacting relativistic
Bose gas have been investigated in numerous studies.
Ref.~\cite{bernstein-1991} presents a general formalism for
calculating the partition function of a relativistic Bose-Einstein
gas. In Ref.~\cite{toms-1994}, the author examines the Bose-Einstein
condensation of a relativistic charged scalar field in a general
homogeneous magnetic field within a space-time of arbitrary
dimension. Ref.~\cite{shiokawa-1999} investigates Bose-Einstein
condensation of a relativistic ideal Bose gas confined in a cavity.
The authors of Ref.~\cite{marko-2014} discuss Bose-Einstein
condensation of a charged scalar field and the Silver Blaze property
within the two-loop $\Phi$-derivable approximation. A series of
papers explored fluctuations in a relativistic ideal Bose gas of
charged particles using various theoretical frameworks; see
Refs.~\cite{zozulya-2005,begun-2007,begun-2008} and references
therein. A comprehensive study of meson condensation can be found in
\cite{mannarelli-2019}.
The possibility of high-temperature BEC in ultrarelativistic nucleus
collisions has been investigated
in Refs.~\cite{begun-2007,begun-2008,begun-2015,deb-2021}.

The present work is organized as follows:
Section \ref{sec:lagrangian-ideal} provides a brief
description of the bosonic system of non-interacting particles and
antiparticles within the framework of the scalar field, considering the
occurrence of BEC.
In Section \ref{sec:bosonic-gas}, the thermodynamic properties of an ideal
Bose gas
comprising particles and antiparticles with conserved charge (isospin) are
examined using the Canonical Ensemble formalism.
This is achieved by replacing the variables of the Grand Canonical Ensemble
with those inherent to the Canonical Ensemble through Legendre transformation.
It is demonstrated that at a non-zero value of the charge (isospin),
a second-order phase transition occurs in the system.
In Section \ref{sec:symmetry-breaking}, we demonstrate that in the bosonic
system, the symmetry breaking of the ground state at $T = 0$  is a consequence
of a first-order phase transition.
The concluding remarks are given in Section \ref{sec:conclusions}.

\section{Free bosonic system of particles and antiparticles:
The scalar field description}
\label{sec:lagrangian-ideal}

We start from the Lagrangian density for the free complex scalar
field\footnote{Here and below we adopt the system of units $\hbar = c = 1$.}
\begin{equation}
\mathcal{L}(x) \,=\, \partial_\mu \hat \phi^+(x) \, \partial^\mu \hat \phi(x)
- m^2\, \hat \phi^+(x) \hat \phi(x)  \,,
\label{eq:lagrangian-ideal-1}
\end{equation}
where $x = (t, \bs r)$ and $\partial_\mu \partial^\mu  =
\partial^2/\partial t^2 - \bs \nabla^2 $.
The action corresponding to the scalar system is given by
\begin{equation}
S[\hat\phi,\hat\phi^+] \,=\, \int d^4x \, \mathcal{L}(x) \,.
\label{eq:action-ideal-1}
\end{equation}
Minimizing the action while accounting for the Lagrangian density
(\ref{eq:lagrangian-ideal-1}) yields the equations of motion for the free fields
$\phi(x)$ and $\phi^+(x)$:
\begin{eqnarray}
\label{eq:eom-ideal-1}
\partial_\mu \partial^\mu \hat \phi(x) \,+\, m^2\, \hat \phi(x) \, &=& 0 \,,
\\
\partial_\mu \partial^\mu \hat \phi^+(x) \,+\, m^2\, \hat \phi^+(x) &=& 0 \,.
\label{eq:eom-ideal-2}
\end{eqnarray}
In the presence of a Bose-Einstein condensate in the system, the scalar field
decomposition, as described by Bogolyubov \cite{bogolyubov}, takes the form
\begin{equation}
\hat\phi(x) \,=\, \Phi_0 + \hat \psi(x) \,,
\label{eq:decompos}
\end{equation}
where $\langle \hat \phi(x) \rangle = \Phi_0$ and $\langle \hat \psi(x) \rangle = 0$.
The field operators $\hat \phi(x)$ and
$\hat \pi(x) = \partial \mathcal{L}/\partial (\partial_0 \hat \phi)
= \partial \hat \phi^+(x)/\partial x^0$
satisfy the following equal-time commutation relations
\begin{equation}
\left[\hat \phi(t,\bs r),\, \hat \pi(t,\bs r')\right]\,
=\, i \delta^3(\bs r - \bs r') \,.
\label{eq:com-rel-10}
\end{equation}
Let us insert the scalar field taking in the form (\ref{eq:decompos})
into the commutation relations
(\ref{eq:com-rel-10}), and take into account that $\hat \phi$ and $\hat \psi$ are
the field operators, but $\Phi_0$ is a wave function,
\begin{eqnarray}
\left[\Phi_0 + \hat \psi(t,\bs r),\,
\frac{\partial \left(\Phi_0^+(t) + \hat \psi^+(t,\bs r')\right)}{\partial t}\right]\,
=\, \left[\hat \psi(t,\bs r),\, \frac{\partial \hat \psi^+(t,\bs r')}{\partial t}\right]
=\, i \delta^3(\bs r - \bs r')\,.
\label{eq:com-rel-20}
\end{eqnarray}
Hence, the thermal field $\hat \psi(x)$ satisfies the same equal-time
commutation relations as the total scalar field $\hat \phi(x)$.
Using this, we can express Bogolyubov's decomposition in the relativistic case as
\begin{eqnarray}
\hat \phi(t,{\bf r}) \,=\,
\Phi_0(t) \,+\,
 \int_{\bs p \neq 0} \frac{d^3p}{(2\pi)^3 2\omega_p} \left[ a_{\bs p}e^{-i \omega_p t}
 + b^+_{\bs p}e^{i \omega_p t}\right] e^{i{\bf p}\cdot {\bf r}} \,,
 \label{eq:expansion-phi-ideal-1}
 \end{eqnarray}
where $\omega_p = \sqrt{m^2 + {\bs p}^2}$ and we assume that the condensate
medium is spatially homogeneous.
Then, in accordance with equal-time commutation relations (\ref{eq:com-rel-20}),
the boson annihilation and creation operators,
$a_{\bs k}$, $a^+_{\bs k}$ and $b_{\bs k}$, $b^+_{\bs k}$,
satisfy the commutation relations
\begin{equation}
\big[ \ a_{\bs k},\, a^+_{\bs k'} \big] \,=\, (2\pi)^3\,2\omega_k \,
\delta^3(\bs k - \bs k') \,,
\qquad
\big[ b_{\bs k},\, b^+_{\bs k'} \big] \,=\, (2\pi)^3\,2\omega_k \,
\delta^3(\bs k - \bs k') \,.
\label{eq:com-rel-ideal}
\end{equation}

After a decomposition of the scalar field (\ref{eq:decompos}) the Lagrangian
density (\ref{eq:lagrangian-ideal-1}) looks like
\begin{eqnarray}
\mathcal{L}(t,\bs r) &=&
\left[\partial_\mu \Phi_0^+ \, \partial^\mu \Phi_0 - m^2 \Phi_0^+\Phi_0\right]
+ \left[\partial_\mu \hat\psi^+(x) \, \partial^\mu \hat\psi(x) - m^2\,
\hat\psi^+(x) \hat\psi(x)\right] +
\nonumber \\
&& + \partial_\mu \Phi_0^+ \, \partial^\mu \hat\psi(x)
+ \partial_\mu \hat\psi^+(x) \,
\partial^\mu \Phi_0 - m^2 \big(\Phi_0^+ \hat\psi + \Phi_0 \hat\psi^+ \big) \,.
\label{eq:lagrangian-ideal-2}
\end{eqnarray}
Let us integrate the field $\hat\phi(t,{\bf r})$ given in
eq.~(\ref{eq:expansion-phi-ideal-1}) over the spatial variables with account
for $V \to \infty$, where $V$ is the total volume of the system.
As a result we get
\begin{eqnarray}
\int d^3r \, \hat\phi(t,{\bf r}) &=& \int d^3r \, \Phi_0 \,
+\, \int_{\bs p \neq 0} \frac{d^3p}{2\omega_p} \left[ a_{\bs p}e^{-i \omega_p t}
+ b^+_{\bs p}e^{i \omega_p t}\right] \delta({\bf p})
\nonumber \\
&=&  V \Phi_0(t) \,,
\label{eq:expansion-phi-ideal-2}
\end{eqnarray}
where we use that
$\int d^3r \exp{(i {\bf p} \cdot {\bf r})} = (2\pi)^3 \delta({\bf p})$.
Therefore, the second integral on the r.h.s. of this equation is zero because the
zero value of the momentum, i.e. $\bs p = 0$, is not included to the integration
interval.
In what follows we investigate the bosonic systems in thermodynamic limit
$V \to \infty$, that is instructive to fix the following relations
\begin{equation}
\int d^3r \, \hat\psi(t, \bs r) \,=\, 0 \,,  \quad  \quad
\int d^3r \, \partial^\mu \hat\psi(t, \bs r) \,=\, 0 \,.
\label{eq:zero-int-psi}
\end{equation}
With account to these relations we integrate the Lagrangian density
(\ref{eq:lagrangian-ideal-2}) over the spatial volume occupied by the system,
$L(t) = \int d^3r \mathcal{L}(t,\bs r)$, and obtain
\begin{eqnarray}
L(t) \,=\,  V \left[\partial_t \Phi_0^+ \partial_t \Phi_0 - m^2 \Phi_0^+\Phi_0\right]
+ \int d^3r \left[\partial_\mu \hat\psi^+(x) \partial^\mu \hat\psi(x)
- m^2\hat\psi^+(x) \hat\psi(x)\right] \,.
\label{eq:lagrangian-ideal-3}
\end{eqnarray}
Here, all terms in eq.~(\ref{eq:lagrangian-ideal-2}) that are linear on the
thermal scalar field $\hat\psi$ disappear after integration over the spatial
volume in accordance with eqs.~(\ref{eq:zero-int-psi}).
However, it will be not true in case if the scalar condensate field $\Phi_0$
depends on spatial coordinates.

Using Lagrangian (\ref{eq:lagrangian-ideal-3}) one can write the action, which
is a functional of two fields:
\begin{eqnarray}
S[\Phi_0,\hat\psi] =  \int dt \, L(t) \,.
\label{eq:action-ideal-2}
\end{eqnarray}
Search for extremum of the action (\ref{eq:action-ideal-2}) results in two
equations of motion for the condensate field $\Phi_0(t)$ and thermal field
$\hat\psi(x)$:
\begin{eqnarray}
\label{eq:eq-phi0-ideal}
\left(\partial_t^2 + m^2 \right) \Phi_0(t) &=& 0 \,,
\\
\left(\partial_\mu \partial^\mu + m^2\right) \hat\psi(x) &=& 0 \,.
\label{eq:eq-psi-ideal}
\end{eqnarray}
There are two solutions of eq.~(\ref{eq:eq-phi0-ideal}) for the condensate field.
First solution is trivial
\begin{equation}
\Phi_0 \,=\, 0 \,,
\label{eq:solution-phi-zero}
\end{equation}
what means that there is no condensate in the system for the total temperature
interval, starting from zero temperature $T = 0$.
In other words, condensate is not formed, the symmetry of the ground state of
the system is not broken, and/or the phase transition to the condensate phase
is absent.

The second solution corresponds to the presence of a spatially homogeneous
condensate in the system
\begin{eqnarray}
\Phi_0(t) \,=\, \frac{1}{\sqrt{2mV}} \left( a_0 e^{-imt} \,+\, b^+_0 e^{imt}\right)
\,=\,  \frac{e^{-imt}}{\sqrt{2m}} \, \psi_0^{(-)}
+ \frac{e^{imt}}{\sqrt{2m}} \, \left(\psi_0^{(+)}\right)^* \,,
\label{eq:phi0-ideal}
\end{eqnarray}
where we introduce the notation:
$a_0/\sqrt{V} = \psi_0^{(-)}$ and $b_0/\sqrt{V} = \psi_0^{(+)}$.
With this notation, we emphasize that $\psi_0^{(-)}$ and $\psi_0^{(+)}$ are
nonrelativistic wave functions associated with particles and antiparticles
at rest.
Concerning eq.~(\ref{eq:eq-psi-ideal}), we now see that the expansion
(\ref{eq:expansion-phi-ideal-1}) of the thermal field $\hat\psi(x)$ can be
interpreted as an expansion in terms of on-shell solutions of the Klein-Gordon
equation (\ref{eq:eq-psi-ideal}).

Taking into account the Lagrangian (\ref{eq:lagrangian-ideal-3}), the Hamiltonian of
the ideal scalar system, $H = \int d^3r \mathcal{H}(\bs r)$, takes the form
\begin{eqnarray}
H = V \left(\Pi_0 \Pi_0^+ + m^2 \Phi_0^+ \Phi_0\right)
+ \int d^3r  \left[\hat\pi(x) \hat\pi^+(x)
+ \bs \nabla \hat\psi^+(x) \cdot \bs \nabla \hat\psi(x)
+ m^2 \hat\psi^+(x) \hat\psi(x)\right] ,
\label{eq:hamiltonian-ideal-1}
\end{eqnarray}
where $\Pi_0 = \partial_0 \Phi_0^+$ and $\hat\pi = \partial_0 \hat\psi^+$.
With the help of solution (\ref{eq:phi0-ideal}) one calculates the condensate
contribution to the effective Hamiltonian that is the first brackets
on the r.h.s. of eq.~(\ref{eq:hamiltonian-ideal-1}), it reads
 \begin{equation}
V \left(\partial_t \Phi_0^+ \, \partial_t \Phi_0 + m^2 \Phi_0^+ \Phi_0 \right)\,
 =\, V m n_{\rm cond} \,,
 \label{eq:cond-contrib-1}
 \end{equation}
where we introduce notation for the total condensate number density
$n_{\rm cond}$:
\begin{equation}
n_{\rm cond} \,=\, \left(\psi_0^{(-)}\right)^*\psi_0^{(-)} \, +\,
\left(\psi_0^{(+)}\right)^*\psi_0^{(+)} \,,
\label{eq:cond-contrib-2}
\end{equation}
Here we use that $\psi_0^{(\pm)}$ are the nonrelativistic wave functions.
With taking this into account we can write the Hamiltonian
(\ref{eq:hamiltonian-ideal-1}) of the ideal bosonic system as
\begin{eqnarray}
H \,=\, m N_{\rm cond}
+ \int d^3r \left[\partial_0 \hat\psi^+(x) \partial_0 \hat\psi(x)
+ \bs \nabla \hat\psi^+(x) \cdot \bs \nabla \hat\psi(x)
+ m^2 \hat\psi^+(x) \hat\psi(x)\right] \,,
\label{eq:hamiltonian-ideal-2}
\end{eqnarray}
where $N_{\rm cond} = V n_{\rm cond}$.
It is obvious that if there is no condensate, the Hamiltonian is determined only
by the thermal states $\hat\psi(x)$.
Using expansion (\ref{eq:expansion-phi-ideal-1}) one can represent the thermal
Hamiltonian $H_{\rm th}$ in terms of the creation and annihilation operators
(see \cite{itzykson-1983,peskin-schroeder-1995}):
\begin{equation}
H \,=\, m N_{\rm cond} \,+\, H_{\rm th} \,,
\qquad
H_{\rm th} \,=\, \frac 12 \int \frac{d^3k}{(2\pi )^3}\,
\left( a^+_{\bs k}\, a_{\bs k}\, +\, b^+_{\bs k}\, b_{\bs k} \right) \,,
\label{eq:hamiltonian-complex}
\end{equation}
where a pure vacuum contribution is not written.

\section{Bosonic ideal gas of particles and antiparticles within
the Extended Canonical Ensemble}
\label{sec:bosonic-gas}

\subsection{Legendre Transformation: Moving from Grand Canonical
Ensemble to Canonical Ensemble in Thermal Phase (isospin conservation) }
\label{sec:legendre-1}

We will study an ideal gas of particles and antiparticles, where the exact
conservation of isospin (charge) is maintained.
Support for a precise conservation law will be conducted within the
extended Canonical Ensemble.
To make this extension, we first formulate the problem in
the Grand Canonical Ensemble (GCE), where the canonical variables are
$(T,\mu_I,V)$.
Then, we make the Legendre transformation and transform the problem to the
framework of the Extended Canonical Ensemble (ECE), where the canonical
variables are $(T,N_I,V)$, whereas the chemical potential becomes the
thermodynamic function, $\mu_I = \mu_I(T,n_I)$.
These steps give us a possibility to take into account the conservation of
the charge in an explicit form.
(In the Appendix \ref{sec:legendre-transform-single-component} we present
these steps: GCE + Legendre transformation $\, \to \,$ ECE,
in the case of a single-component ideal bosonic gas.
We compare this approach to the method that explicitly includes the
Kronecker delta symbol in the partition function, ensuring the exact
conservation of $N_I$.)

So, to describe the thermodynamics of the system, we begin by analyzing it
within the Grand Canonical Ensemble.
Initially, this description focuses solely on the thermal phase, meaning we
consider $T > T_{\rm c}$.
The Hamiltonian (\ref{eq:hamiltonian-complex}) that contains only
thermal degrees of freedom reads
\begin{equation}
H \,=\, \int \frac{d^3k}{(2\pi)^3 2\omega_k} \,
\omega_k \, \left[ a^+_{\bs k} a_{\bs k} + b^+_{\bs k} b_{\bs k}\right]\,.
\label{eq:id2-hamiltonian-therm}
\end{equation}
From equation of motion for the free field (\ref{eq:eom-ideal-1})
- (\ref{eq:eom-ideal-2}) follows the continuity equation
$\partial_\mu j^{\mu}(x) = \partial j_0/\partial t - \bs \nabla \bs j =0$,
which we also understood as the Noether current for charge conservation.
The current of the free fields in the operator form reads
\begin{equation}
\hat j^{\mu}(x)\, =\, i \left[\hat\phi^\dagger(x) \partial^\mu \hat\phi(x) \,
-\, \left(\partial^\mu \hat\phi^\dagger(x) \right)\, \hat\phi(x)   \right] \,.
\label{eq:free-current}
\end{equation}
The isospin (charge) density operator is the time component of the current
\begin{eqnarray}
\hat n_I(x) \, \equiv \,  \hat j^0(x) \,
=\, i \left[ \hat\phi^\dagger(x) \frac{\partial \hat\phi(x)}{\partial x_0} \,
-\, \frac{\partial \hat\phi^\dagger(x)}{\partial x_0} \,\hat\phi(x)   \right] \,.
\label{eq:operator-charge-density-1}
\end{eqnarray}
Operator $\hat N_I$ for the total number of thermal particles is obtained as
an integral over the total volume of the system of the isospin (charge)
density (\ref{eq:operator-charge-density-1}):
\begin{equation}
\hat N_I\,=\, i \int d^3r \,
\left[\hat\psi^\dagger(x) \partial_t \hat\psi(x) \,
-\, \left(\partial_t \hat\psi^\dagger(x) \right)\, \hat\psi(x)   \right] \,.
\label{eq:oper-total-number}
\end{equation}
Then, in the representation of the annihilation and creation operators the
operator $\hat N_I$ for thermal particles is given by
%
\begin{eqnarray}
\hat N_I \,=\, \int \frac{d^3k}{(2\pi)^3 2\omega_k}\,
\left[ a^+_{\bs k} a_{\bs k} -  b^+_{\bs k} b_{\bs k}\right] \,.
\label{eq:id2-isospin-operator}
\end{eqnarray}
Using the Hamiltonian (\ref{eq:id2-hamiltonian-therm}) and the isospin number
operator (\ref{eq:id2-isospin-operator}), we calculate the partition function
of the Grand Canonical Ensemble
\begin{equation}
Z(T,\mu_I,V) \,=\, {\rm Tr} \exp{\left[-( H - \mu_I \hat N_I)/T \right]} \,.
\label{eq:charge-part-func-id-gas}
\end{equation}
The thermodynamic potential $\Omega(T,\mu_I,V) = - T \ln Z(T,\mu_I,V)$,
after explicit calculation of the partition function
(\ref{eq:charge-part-func-id-gas}) has the form
\begin{equation}
\Omega(T,\mu_I,V) \,=\, T V \int \frac{d^3k}{(2\pi)^3} \left\{
\ln{ \left[ 1- e^{- \big(\omega_k - \mu_I \big)/T } \right] }
+ \ln{ \left[ 1- e^{- \big(\omega_k + \mu_I \big)/T } \right] } \right\} \,.
\label{eq:charge-omega-id}
\end{equation}
In the Grand Canonical Ensemble the isospin number $N_I$ is connected with
correspondent chemical potential $\mu_I$ as:
\begin{equation}
N_I \,=\, - \, \frac{\partial \Omega}{\partial \mu_I} \,.
\label{eq:charge-legendre-transform-1}
\end{equation}
We will use this relation as a basis for the Legendre transformation.
With the help of expression (\ref{eq:charge-omega-id}), one can rewrite
eq.~(\ref{eq:charge-legendre-transform-1})
\begin{eqnarray}
N_I \,=\, V\, \int \frac{d^3k}{(2\pi)^3}\,
\left[f_{\rm BE}(\omega_k, \mu_I) - f_{\rm BE}(\omega_k, - \mu_I)\right] \,,
\label{eq:id2-therm-charge-density-eq}
\end{eqnarray}
where the Bose-Einstein distribution function reads
\begin{eqnarray}
f_{\rm BE}(\omega_k,\mu_I)=
\left[\exp{\left(\frac{\omega_k - \mu_I}{T}\right)} - 1\right]^{-1}
\quad {\rm with} \quad \omega_k = \sqrt{m^2+\bs k^2} \,.
\end{eqnarray}
Solving eq.~(\ref{eq:id2-therm-charge-density-eq}) with respect to $\mu_I$
we get dependence of the chemical potential on temperature $T$ and isospin
number $N_I$,  $\mu_I = \mu_I(T,N_I/V)$.
Next, we define the free energy $F$ as:
\begin{equation}
F(T,N_I,V) \,=\, \Omega \,-\, \mu_I \, \frac{\partial \Omega}{\partial \mu_I} \,,
\label{eq:charge-legendre-transform-2}
\end{equation}
where it is implied that the chemical potential $\mu_I$ is a function of
variables $(T,N_I/V)$.
In explicit form this equation reads
\begin{equation}
F(T,N_I,V) \,=\, -T \, \ln{\left\{{\rm Tr} \,
\left[e^{\dis -(H - \mu_I(T,n_I) \hat N_I)/T}\right]\right\}} \,
+\, \mu(T,n_I) \, N_I \,,
\label{eq:legendre-transform-2.2}
\end{equation}
where $n_I = N_I/V$.
With the help of expressions~(\ref{eq:charge-legendre-transform-1})
and (\ref{eq:charge-legendre-transform-2}) one can check that
\begin{equation}
\mu_I \,=\, \frac{\partial F(T,N_I,V)}{\partial N_I} \,.
\label{eq:charge-legendre-transform-5}
\end{equation}
And using connection $\Omega = - p(T,\mu_I) V$ one can obtain that for the
quantity $F$ introduced in (\ref{eq:charge-legendre-transform-2}) it is valid
 $\Phi(T,n_I) = - p(T,\mu_I(T,n_I)) + \mu_I n_I$, where $\Phi = F/V$.
This relation together with eq.~(\ref{eq:charge-legendre-transform-5})
confirm that introduced $F$ coincides indeed with the thermodynamic free energy.

Using in eq.~(\ref{eq:charge-legendre-transform-2}) the explicit form of the
thermodynamic potential (\ref{eq:charge-omega-id}), one can write down the free
energy for an ideal gas of particles and antiparticles in the case of
conservation of charge
\begin{equation}
\Phi(T,n_I) \,=\, \mu_I \, n_I \,+\, T \int \frac{d^3k}{(2\pi)^3} \left\{
\ln{ \left[ 1- e^{- \big(\omega_k - \mu_I \big)/T } \right] }
+ \ln{ \left[ 1- e^{- \big(\omega_k + \mu_I \big)/T } \right] } \right\} \,,
\label{eq:charge-legendre-transform-id}
\end{equation}
where the chemical potential is a thermodynamic function $\mu_I(T,n_I)$.
From the latter expression follows that the pressure
$p(T,n_I) = - \Phi(T,n_I) +  n_I \mu_I(T,n_I)$
in an ideal particle-antiparticle boson gas looks like
\begin{eqnarray}
p(T,n_I) &=& - T \int \frac{d^3k}{(2\pi)^3} \, \left\{
\ln{ \left[ 1- e^{- \big(\omega_k - \mu_I \big)/T } \right] }
+ \ln{ \left[ 1- e^{- \big(\omega_k + \mu_I \big)/T } \right] }
\right\}
\nonumber \\
&=& \frac 13 \int \frac{d^3k}{(2\pi)^3} \, \frac{{\bf k}^2}{\omega_k} \,
\left[f_{\rm BE}(\omega_k, \mu_I) + f_{\rm BE}(\omega_k, - \mu_I)\right] \,,
\label{eq:id2-pressure-term}
\end{eqnarray}
where we recall once again that the chemical potential is a thermodynamic
function, $\mu_I(T,n_I)$.

So, by performing the Legendre transformation, we have made the transition from
the Grand Canonical Ensemble, which deals with the free variables $(T,\mu_I)$,
to the Extended Canonical Ensemble (ECE), which deals with the free
variables $(T,n_I)$.
So, now, using ECE, we can consider problems where the isospin density
is taking into account as a given quantity, i.e., explicitly establishing
that $n_I =$~const.

Averaging the Hamiltonian (\ref{eq:id2-hamiltonian-therm}) we calculate
the total energy of the system
\begin{equation}
E = \langle H \rangle \,=\, \int \frac{d^3k}{(2\pi)^3 2\omega_k} \,
\omega_k \, \left[\langle a^+_{\bs k} a_{\bs k} \rangle \,
+\, \langle b^+_{\bs k} b_{\bs k} \rangle \right] \,.
\label{eq:id2-energy-therm}
\end{equation}
The statistical mean values of the particle-number operators are taken as
\cite{itzykson-1983}:
\begin{eqnarray}
\label{eq:mean-on-app5a}
\big\langle a^+_{\bs k} a_{\bs k} \big\rangle
\,=\, V\, 2\omega_k\, f_{\rm BE}(\omega_k, \mu_I) \,,
\qquad
\big\langle b^+_{\bs k} b_{\bs k} \big\rangle
\,=\, V\, 2\omega_k\, f_{\rm BE}(\omega_k, - \mu_I) \,.
\end{eqnarray}
Using these mean occupation numbers of the thermal particles
for the mean energy density $\varepsilon = E/V$ we get
\begin{equation}
\varepsilon \,=\, \int \frac{d^3k}{(2\pi)^3} \, \omega_k \,
\left[f_{\rm BE}(\omega_k, \mu_I) + f_{\rm BE}(\omega_k, - \mu_I)\right] \,.
\label{eq:id2-energy-dens-therm}
\end{equation}
The entropy density $s(T,n_I)$ is calculated using the free energy
(\ref{eq:charge-legendre-transform-id}):
$s = [ - \partial \Phi(T,n_I)/\partial T ]_{n_I}$.
As a result we obtain
\begin{equation}
T s = p(T,n_I) \,+\, \varepsilon(T,n_I) \,-\, n_I \, \mu_I(T,n_I) \,.
\label{eq:ideal2-entropy-2}
\end{equation}
Hence, we obtain the right thermodynamic relation between densities of the main
thermodynamic quantities, i.e. the Euler relation.

\medskip

\noindent
{\it The Stefan-Boltzmann limit.} \\
It is instructive to calculate the Stefan-Boltzmann limit,
i.e. the energy density at high temperatures
\begin{equation}
\lim_{T \to \infty} \frac{\varepsilon}{T^4}
\approx 2\!\!\int\frac{d^3k}{(2\pi)^3}\frac{\sqrt{m^2+\bs k^2}}
{e^{\sqrt{m^2+\bs k^2}/T}-1}
= \frac{\tilde m^4}{\pi^2} \int_0^\infty dx x^2 \sqrt{1 + x^2}
\frac{e^{-\tilde m \sqrt{1 + x^2}}}{1 - e^{-\tilde m \sqrt{1 + x^2}}} \,,
\label{eq:id2-sb-1a}
\end{equation}
where $x = |\bs k|/m$, $\tilde{m} = m/T$,
and we neglect the ratio $\lim_{T \to \infty} \mu_I/T \to 0$.
Next we take that $\exp{(-\tilde m \sqrt{1 + x^2})} < 1$ and make an expansion
$[1 - \exp{(-\tilde m \sqrt{1 + x^2})}]^{-1}
= 1 + \sum_{n = 1}^{\infty}  \exp{(-n \tilde m \sqrt{1 + x^2})}$.
%
%
To calculate integral we use that for some $\alpha < 1$ there is approximation
$\int_0^\infty dx  x^2  \sqrt{1 + x^2} \exp{(-\alpha \sqrt{1 + x^2})}
\approx 6/\alpha^4$.
%
%
%
Hence
\begin{eqnarray}
\lim_{T \to \infty} \frac{\varepsilon}{T^4} =
\frac{\tilde m^4}{\pi^2} \int_0^\infty dx x^2 \sqrt{1 + x^2} \,
\sum_{n = 1}^{\infty} \, e^{-n \tilde m \sqrt{1 + x^2}}
\approx \, \frac{6}{\pi^2} \sum_{n = 1}^{\infty} \, \frac{1}{n^4} \,
=\, \frac{6}{\pi^2} \, \zeta(4) \,,
\label{eq:id2-sb-4}
\end{eqnarray}
where $\zeta(4) = \pi^4/90$ is the Reman zeta-function.
So, finally we obtain
\begin{equation}
\lim_{T \to \infty} \frac{\varepsilon}{T^4} \,=\, \frac{\pi^2}{15} \,.
\label{eq:id2-sb-5}
\end{equation}
In fact, it is seen in Fig.~\ref{fig:ideal-2comp-gas-phys-quantitie} on the
left panel that the energy density approaches the value
$\sigma_{_{\rm SB}} = \pi^2/15$ that is the Stefan-Boltzmann constant for
a two-component boson system.

\medskip

\noindent
{\it One comment to conclude this section}

It can be argued that within the Canonical Ensemble, there is a well-known
thermodynamic relation $dF =-s dT -p dV + \mu_I dN_I$ for describing a system
with charge (isospin) variations.
Therefore, the chemical potential can be defined using equation
(\ref{eq:charge-legendre-transform-5}).
Then, there seems to be no need for the scheme proposed above.
However, it should be noted that eq.~(\ref{eq:charge-legendre-transform-5})
is a thermodynamic definition of the chemical potential; it does not provide
a recipe for using the chemical potential in statistical microscopic
calculations, nor a method for incorporating the chemical potential into
description of the system when the isospin is kept constant,
$N_I =$~const, across different temperatures.

At the same time, the method proposed in the present paper provides a formally
correct scheme for incorporating the chemical potential into the system's
description {\it at the microscopic level} within the Canonical Ensemble
framework.
Indeed, the thermodynamic properties of the bosonic system are analyzed
on base of the quantum statistical averaging giving in
eq.~(\ref{eq:legendre-transform-2.2}); that approach can be named as
the Extended Canonical Ensemble (ECE).
In ECE, the average values are now obtained using the following
statistical operator
\begin{equation}
\rho(T,n_I) \,
=\, \exp{\left[- \frac 1T \left(H - \mu_I(T,n_I) \hat N_I\right)\right]} \,,
\label{eq:ece-stat-oper-charge}
\end{equation}
which is defined for a given temperature $T$ and a given charge (isospin)
density $n_I$.
The partition function, calculated using the extended statistical operator
(\ref{eq:ece-stat-oper-charge}), along with an understanding of the chemical
potential as a function of temperature and isospin density, paves
the way for the microscopic calculations of all thermodynamic quantities.
In particular, this approach allows us to consider problems in which the
isospin number density is exactly kept constant, $n_I =$~const, at different
temperatures.

\subsection{Legendre Transformation: Moving from Grand Canonical Ensemble
to Canonical Ensemble in Condensate Phase (isospin conservation) }
\label{sec:legendre-2}

We are going to consider a bosonic system of particles and antiparticles
in the condensate phase, i.e., in the temperature interval $T \le T_{\rm c}$,
in the case of isospin conservation.
As was obtained in (\ref{eq:hamiltonian-complex}) in the condensate phase
the Hamiltonian
in addition to the thermal degrees of freedom contains also energy of the
condensed particles
\begin{equation}
H \,=\, \varepsilon_0 N_{\rm cond} + \frac 12 \int \frac{d^3k}{(2\pi)^3} \,
\left[ a^+_{\bs k} a_{\bs k} + b^+_{\bs k} b_{\bs k} \right]\,.
\label{eq:charge-id-hamiltonian-cond}
\end{equation}
 where $\varepsilon_0 = m$ is the single-particle ground state.
The operator $\hat N_I$ of the total isospin number of the system is
obtained, as an integral of the zero-component $\hat j^0$ of the current
(\ref{eq:free-current}), over the volume of the system:
\begin{equation}
\hat N_I(t,\bs r)\,=\, i \int d^3r \,
\left[\hat\phi^\dagger(t,\bs r) \partial_t \hat\phi(t,\bs r) \,
-\, \left(\partial_t \hat\phi^\dagger(t,\bs r) \right)\, \hat\phi(t,\bs r) \right] \,.
\label{eq:integral-zero-current}
\end{equation}
Using eqs.~(\ref{eq:expansion-phi-ideal-1}) and (\ref{eq:zero-int-psi}) one can
show that in a homogeneous system the isospin-number (charge-number)
operator $\hat N_I$ is separated into two contributions
\begin{eqnarray}
\hat N_I \,=\, N_{I\rm cond} +  \hat N_{I\rm th}
\qquad  {\rm with}  \qquad
\hat N_{I\rm th} \,=\, \int \frac{d^3k}{(2\pi)^3 2\omega_k}\,
\left[ a^+_{\bs k} a_{\bs k} - b^+_{\bs k} b_{\bs k}\right] \,,
\label{eq:charge-number-id}
\end{eqnarray}
where we made a regularization of the integral over momentum.
Here $N_{I\rm cond} = V n_{I_{\rm cond}}$, where with account
for eq.~(\ref{eq:phi0-ideal}) we get
\begin{equation}
n_{I_{\rm cond}} \,=\, i \left[ \Phi_0^+\, \partial_t \Phi_0\,
-\, \left(\partial_t \Phi_0^+\right)\, \Phi_0 \,\right] \,
=\, \left(\psi_0^{(-)}\right)^*\psi_0^{(-)} \,
-\, \left(\psi_0^{(+)}\right)^*\psi_0^{(+)} \,.
\label{eq:cond-contrib-11}
\end{equation}
It is instructive to compare this result with eq.~(\ref{eq:cond-contrib-2}),
where we define the total density of the particles that are in the ground state.
Using two expressions (\ref{eq:charge-id-hamiltonian-cond}) and
(\ref{eq:charge-number-id}) within the GCE one calculates the partition
function in the temperature interval $T \le T_{\rm c}$:
\begin{equation}
Z(T,\mu_I,V) \,=\,
e^{- \beta \left(\varepsilon_0 N_{\rm cond} - \mu_I N_{I\rm cond} \right)} \,
{\rm Tr}\left[ e^{- \beta \left[H - \mu_I \hat N_I \right]_{\rm th}}\right] \,,
\label{eq:charge-id-gce-part-func-cond-2}
\end{equation}
where we introduce notation for pure thermal quantities:
\begin{equation}
\left[H - \mu_I \hat N_I \right]_{\rm th}
=  \int \frac{d^3k}{(2\pi)^3 2\omega_k} \,
\left\{ \left[(\omega_k - \mu_I) a^+_{\bs k} a_{\bs k}
+ (\omega_k + \mu_I) b^+_{\bs k} b_{\bs k}\right] \right\}\,.
\label{eq:charge-id-gce-part-func-cond-3}
\end{equation}
Consequently, one can conclude that the C-number contributions to the
Hamiltonian (\ref{eq:charge-id-hamiltonian-cond}) and particle-number operator
(\ref{eq:charge-number-id}) associated with the presence of the condensate
do not affect the mean value of an operator $\hat A$ in comparison with the
statistical averaging of this operator in the pure thermal state of the system.
Indeed, the mean value reads
\begin{equation}
\left\langle \hat A \right\rangle \,
=\, \frac{1}{Z} \,{\rm Tr}\left[ e^{- \beta \left(H - \mu_I \hat N_I \right)}
\hat A \right] \,
=\ \frac{1}{Z_{\rm th}} \,
{\rm Tr}\left[ e^{- \beta \left(H - \mu_I \hat N_I \right)_{\rm th}}
\hat A \right] \,,
\label{eq:mean-averaging-operator}
\end{equation}
where we use eq.~(\ref{eq:charge-id-gce-part-func-cond-2}) and notation
$Z_{\rm th} =
{\rm Tr} \exp{\left\{ - \beta \left[H - \mu_I \hat N_I \right]_{\rm th}\right\}}$.

Using partition function (\ref{eq:charge-id-gce-part-func-cond-2}) one can
calculate the thermodynamic potential $\Omega = -T \ln Z$:
\begin{eqnarray}
\Omega(T,\mu_I,V) \,=\, \left(\varepsilon_0 N_{\rm cond} - \mu_I N_{I\rm cond} \right)
+\, \Omega_{\rm th}(T,\mu_I,V) \,,
\label{eq:charge-id-omega-cond}
\end{eqnarray}
where the thermal part of the thermodynamic potential reads
\begin{eqnarray}
\Omega_{\rm th}(T,\mu_I,V) \,=\, VT \int \frac{d^3k}{(2\pi)^3} \,
\left\{ \ln{ \left[ 1- e^{- \big(\omega_k - \mu_I \big)/T } \right] }
+ \ln{ \left[ 1- e^{- \big(\omega_k + \mu_I \big)/T } \right] } \right\} \,.
\label{eq:charge-id-omega-cond-1}
\end{eqnarray}
Here and further it is implied that $N_I = N^{(-)} - N^{(+)} \ge 0$,
 $N_{I\rm cond} = N_{\rm cond}^{(-)} - N_{\rm cond}^{(+)}$ with
 $N^{(-)} = N_{\rm cond}^{(-)} + N_{\rm th}^{(-)}$
and  $N^{(+)} = N_{\rm cond}^{(+)} + N_{\rm th}^{(+)}$.

In accordance with eq.~(\ref{eq:charge-id-omega-cond}), in the Grand Canonical
Ensemble the relation between the chemical potential and the isospin number,
$N_I = - \partial \Omega/\partial \mu_I$, reads
\begin{equation}
N_I \,=\, N_{I\rm cond} \,- \, \frac{\partial \Omega_{\rm th}}{\partial \mu_I} \,.
\label{eq:charge-cond-legendre-transform-1}
\end{equation}
We define the free energy in the same way as in previous
Section \ref{sec:legendre-1}:
\begin{equation}
F(T,N_I,V) \,=\, \Omega \,-\, \mu_I \, \frac{\partial \Omega}{\partial \mu_I}\,.
\label{eq:charge-cond-legendre-transform-2}
\end{equation}
Using eq.~(\ref{eq:charge-id-omega-cond}) and the basic derivative
$N_I = - \partial \Omega/\partial \mu_I$ one can rewrite this as
\begin{eqnarray}
F(T,N_I,V) \,=\, \left(\varepsilon_0 N_{\rm cond} - \mu_I N_{I\rm cond}\right)
+ \Omega_{\rm th} \,+\, \mu_I \, N_I \,.
\label{eq:charge-cond-legendre-transform-2a}
\end{eqnarray}
In explicit form eq.~(\ref{eq:charge-cond-legendre-transform-2a}) reads
\begin{eqnarray}
\Phi(T,n_I) &=&  (\varepsilon_0 n_{\rm cond}  - \mu_I n_{I\rm cond}) \,
+\, \mu_I \, n_I \,+
\nonumber \\
&&  +\, T \int \frac{d^3k}{(2\pi)^3} \, \left\{
\ln{ \left[ 1- e^{- \big(\omega_k - \mu_I \big)/T } \right] }
+ \ln{ \left[ 1- e^{- \big(\omega_k + \mu_I \big)/T } \right] } \right\}  \,,
\label{eq:charge-cond-legendre-transform-2c}
\end{eqnarray}
where we define the density of the free energy $\Phi(T,n_I) = F(T,N_I,V)/V$.

Using eqs.~(\ref{eq:charge-cond-legendre-transform-1}) and
(\ref{eq:charge-id-omega-cond-1}) we calculate the isospin number density and
get
\begin{eqnarray}
n_I \,=\, n_{I\rm cond} + \int \frac{d^3k}{(2\pi)^3}\,
\left[f_{\rm BE}(\omega_k, \mu_I) - f_{\rm BE}(\omega_k, - \mu_I) \right] \,,
\label{eq:id-charge-density}
\end{eqnarray}
where $n_{I\rm cond}$ is given in (\ref{eq:cond-contrib-11}).

It is necessary to point out that in ideal gas just one component of the system
can develop a condensate when we keep $n_I \ne 0$.
Indeed, if we assume that both components of the bosonic system develop
condensed states simultaneously, then this means that the following two
relations must be fulfilled  at the same time
\begin{eqnarray}
\label{eq:cond-condensate-p}
m \,-\, \mu_I &=& 0 \,, \quad {\rm particles}
\\
m \,+\, \mu_I &=& 0 \,, \quad {\rm antiparticles}
\label{eq:cond-condensate-antip}
\end{eqnarray}
This leads to the equivalent equalities $m = 0$ and $\mu_I = 0$.
Since we consider the massive boson gas, thus only one of the
eqs.~(\ref{eq:cond-condensate-p}), (\ref{eq:cond-condensate-antip}) is
valid.
This coincides with conclusion that was obtained in
Refs.~\cite{haber-1981,haber-1982,kapusta-1981}.
In our study we adopt that the negatively charged bosons (particles)
represent the charge dominated component of the particle-antiparticle system,
i.e. $N_I = N^{(-)} - N^{(+)} \ge 0$.
Hence, we must adopt that only eq.~(\ref{eq:cond-condensate-p}) is true, or
$\mu_I = m$, and only the negatively charged particles can create condensate.

So, we state that only one component of the system, i.e. particles
in accordance with our choice, is in the condensate phase
(see eq.~(\ref{eq:cond-contrib-2}))
%
\begin{equation}
n_{\rm cond} \,=\, n_{I\rm cond} \,=\, \left(\psi_0^{(-)}\right)^*\psi_0^{(-)} \,,
\label{eq:ncond-min-charge}
\end{equation}
whereas the density of the condensed antiparticles is zero, i.e.
$n_{\rm cond}^{(+)} = \left(\psi_0^{(+)}\right)^*\psi_0^{(+)} = 0$.
Taking that $\mu_I = m$ into account we rewrite eq.~(\ref{eq:id-charge-density}) as
\begin{eqnarray}
n_I \,=\, n_{\rm cond}(T) \,+\, n_{\rm lim}(T) \,-\, \int \frac{d^3k}{(2\pi)^3} \,
f_{\rm BE}(\omega_k, - \mu_I)\big|_{\mu_I = m} \,,
\label{eq:id-charge-density-2}
\end{eqnarray}
where we define the function
\begin{eqnarray}
n_{\rm lim}(T) \,=\,\int \frac{d^3k}{(2\pi)^3}\, f_{\rm BE}(\omega_k,m) \,.
\label{eq:id-cc-density}
\end{eqnarray}
This function represents the maximum or ``limiting'' density of thermal
particles for a given temperature.
According to Einstein's hypothesis published in 1925 \cite{einstein-1925},
all particles exceeding at temperature $T$ the density $n_{\rm lim}(T)$
are concentrated on the ground state $\varepsilon_0$.
Seventy years later, in 1995, this hypothesis was confirmed by the observation
of Bose-Einstein condensation \cite{anderson-1995}.

On the other hand, the line $n_{\rm lim}(T)$ in the $(T,n)$-plane is a
critical curve that separates the condensate phase from the thermal phase.
We see that the curve (\ref{eq:id-cc-density}) coincide with the
critical curve defined for an ideal single-component bosonic gas.
The latter coincidence appears because only one charge-dominant component
can create condensate.
And also for the same reason, all condensate particles are negatively charged.
Remind, we adopt that particles with  negative charge dominate in the system.
In opposite case, the condensate would consist only of positively charged
antiparticles.
Hence, the condensate is homogeneous with respect to charge, i.e. it consists
from only negative or from only positive bosons with zero momentum.

Some features of the definition of free energy
(\ref{eq:charge-cond-legendre-transform-2a}) using Legendre transformation
in the condensate phase are worth noting.
Indeed, in the previous Section \ref{sec:legendre-1}, where only the thermal
phase was considered, the equation $N_I = - \partial \Omega/\partial \mu_I$
was used to obtain dependence $\mu_I = \mu_I(T,n_I)$.
However, in the presence of condensate, the same equation in the form
(\ref{eq:charge-cond-legendre-transform-1}) leads only to a fixed value of
the chemical potential $\mu_I = m$.
Under this condition, the equation that we obtain in the form
(\ref{eq:id-charge-density-2}) is used to obtain the temperature dependence
of the condensate, $n_{\rm cond}(T)$.

Another important conclusion is that the chemical potential takes one of two
values in the condensate phase $\mu_I = \pm m$.
This indicates that the chemical potential cannot change, since the condition
of condensate formation determines its value.
Thus, the chemical potential is no longer a free variable, since it cannot
dictate the thermodynamic state of the system.
Indeed, the system can be characterized by any value $n_I > 0$, but each
of these values corresponds to the same chemical potential $\mu_I = m$.
This means that the Grand Canonical Ensemble is unsuitable for describing
a bosonic system in the condensate phase, and the appropriate approach to
studying a bosonic system must be through the Canonical Ensemble.

The same conclusion applies to a single-component bosonic system, where
a constant particle number density $n$ is maintained.
In the non-relativistic case, when we consider the condensate phase, where
the chemical potential is zero $\mu = 0$, any value of $n$ is permissible.
Thus, the chemical potential cannot determine the thermodynamic state of the
system in the presence of condensate.
Therefore, we cannot study the bosonic system within the Grand Canonical
Ensemble, fixing the variables $(T,\mu)$; however, we can do so using
the Canonical Ensemble, fixing the variables $(T,n)$.

\medskip

\subsubsection{Thermodynamic quantities}

With account for the condensate condition $\mu_I = \varepsilon_0 = m$
one can calculate thermodynamic quantities in the temperature interval
$T \le T_{\rm c}$.
First we rewrite the free energy density
(\ref{eq:charge-cond-legendre-transform-2c}):
\begin{eqnarray}
\Phi(T,n_I) \,=\,  m \, n_I \,
+\, T \int \frac{d^3k}{(2\pi)^3} \, \left\{
\ln{ \left[ 1- e^{- \big(\omega_k - m \big)/T } \right] }
+ \ln{ \left[ 1- e^{- \big(\omega_k + m \big)/T } \right] } \right\}  \,,
\label{eq:cond-dens-free-energy}
\end{eqnarray}
where we use eq.~(\ref{eq:ncond-min-charge}).
With the Hamiltonian (\ref{eq:charge-id-hamiltonian-cond}) we calculate
the energy density
\begin{equation}
E = \langle H \rangle \,=\, \varepsilon_0 N_{\rm cond} \,+\,
\int \frac{d^3k}{(2\pi)^3 2\omega_k} \,
\omega_k \, \left[\langle a^+_{\bs k} a_{\bs k}\rangle \,
+\, \langle b^+_{\bs k} b_{\bs k} \rangle \right] \,.
\label{eq:id-hamiltonian-cond-aver-10}
\end{equation}
As it was shown in (\ref{eq:mean-averaging-operator}) the presence of the
condensate is not affected the averaging of an operator obtained in the
thermal phase.
Therefore, we can use the mean values (\ref{eq:mean-on-app5a}).
Thereby, for the energy density $\varepsilon = E/V$ we get
\begin{equation}
\varepsilon \,=\, \varepsilon_0 n_{\rm cond}(T,n_I) \,
+\, \int \frac{d^3k}{(2\pi)^3} \,
\omega_k \, \left[f_{\rm BE}(\omega_k, \mu_I) + f_{\rm BE}(\omega_k, - \mu_I)
\right]_{\mu_I = \varepsilon_0} \,.
\label{eq:id-hamiltonian-cond-aver-30}
\end{equation}
To complete the equation of state of an ideal particle-antiparticle gas,
we write the pressure in the condensate phase, using the free energy density
(\ref{eq:cond-dens-free-energy}), as $p = m n_I - \Phi$:
\begin{eqnarray}
p(T) &=& - T \int \frac{d^3k}{(2\pi)^3} \, \left\{
\ln{ \left[ 1- e^{- \big(\omega_k - \mu_I \big)/T } \right] }
+ \ln{ \left[ 1- e^{- \big(\omega_k + \mu_I \big)/T } \right] }
\right\}_{\mu_I = \varepsilon_0}
\nonumber \\
&=& \frac 13 \int \frac{d^3k}{(2\pi)^3} \, \frac{{\bf k}^2}{\omega_k} \,
\left[f_{\rm BE}(\omega_k, \mu_I) + f_{\rm BE}(\omega_k, - \mu_I)
\right]_{\mu_I = \varepsilon_0} \,.
\label{eq:charge-id-pressure-cond-2}
\end{eqnarray}
It is necessary to point out, that in the absence of interaction in a system,
the condensate particles do not contribute to kinetic pressure, exactly this
result we obtain in (\ref{eq:charge-id-pressure-cond-2}).

\begin{figure}
\centering
\includegraphics[width=0.495\textwidth]{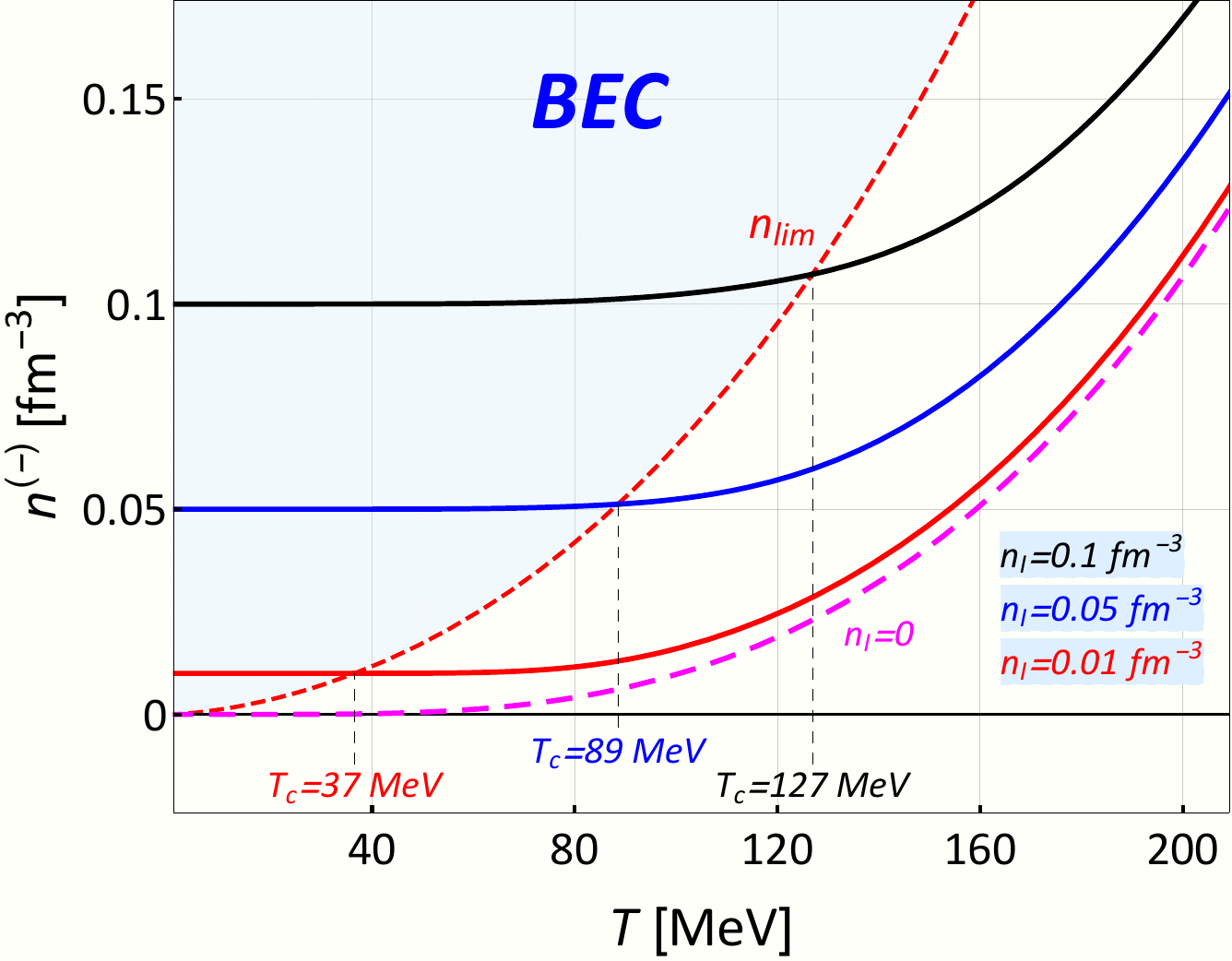}
\includegraphics[width=0.47\textwidth]{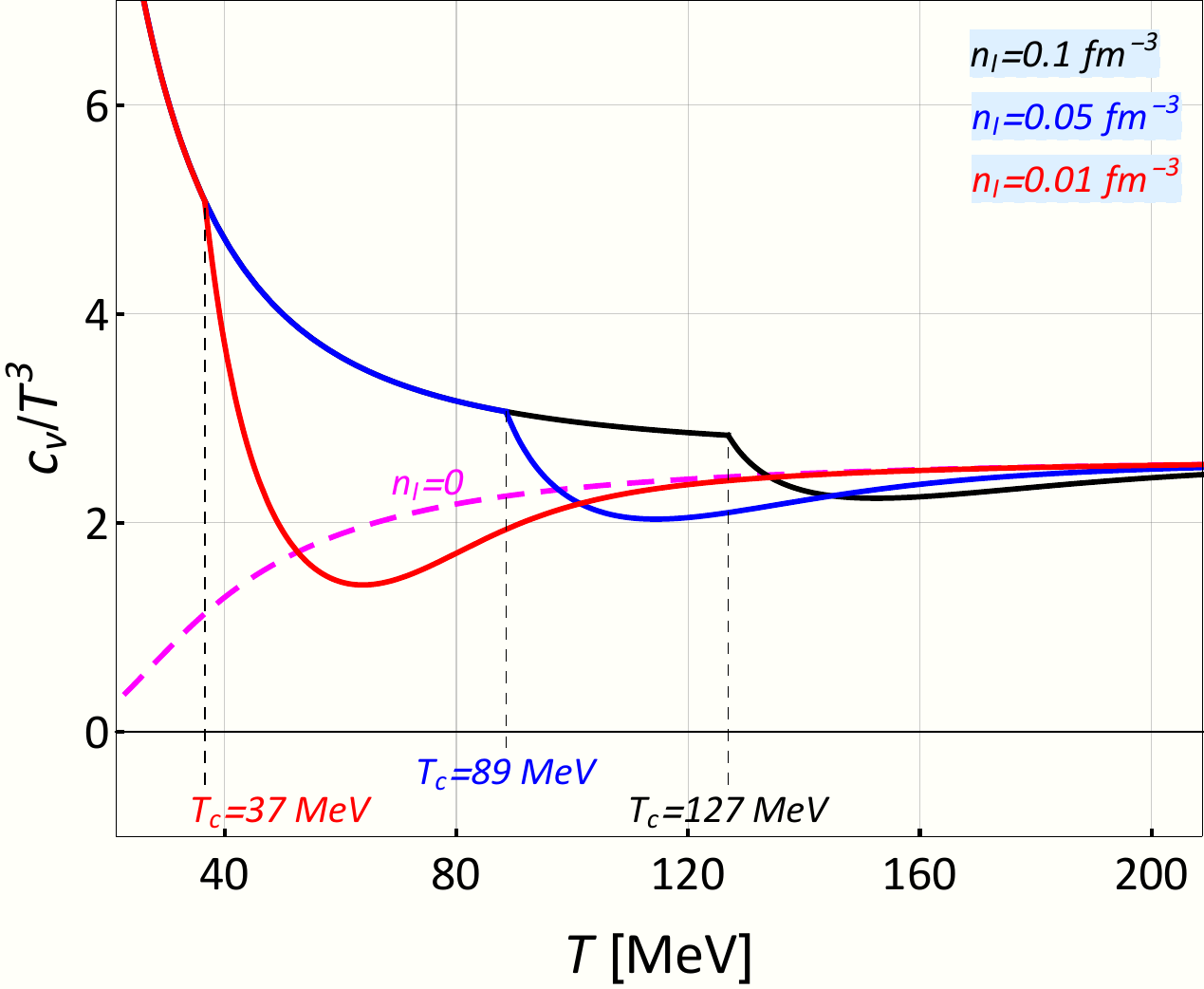}
\caption{An ideal charged bosonic gas of particles and antiparticles ($m = m_\pi$).
{\it Left panel:} Particle density $n^{(-)}$ vs. temperature
at $n_I = 0.1,\, 0.05,\, 0.01$~fm$^{-3}$, red, blue and black
lines respectively, and $n_I = 0$ as the dashed pink line.
The values of $T_{\rm c}$ shown correspond to a second-order phase transition.
{\it Right panel:} Temperature dependence of the heat capacity
with the same parameter values and notation as in the left panel.
 }
\label{fig:ideal-2comp-gas-particle-density}
\end{figure}

In the condensate phase the entropy density
$s(T,n_I) = - \partial \Phi(T,n_I)/\partial T$ is calculated using
the free energy (\ref{eq:cond-dens-free-energy}):
\begin{equation}
T \, s = p(T) \,+\, \varepsilon(T,n_I) \,-\, m \, n_I \,.
\label{eq:ideal2-entropy-4}
\end{equation}
It is interesting to note that after taking derivative $\partial \Phi/\partial T$
we left with only three kinetic contributions.
Meanwhile, if one takes into account that in the condensate phase
the expressions for the isospin and energy densities are the following
\begin{equation}
n_I = n_{\rm cond} + n_{\rm Ith}(T) \,, \qquad
\varepsilon = m n_{\rm cond} + \varepsilon_{\rm th}(T) \,,
\label{eq:replace-ni}
\end{equation}
where $n_{\rm Ith}(T)$ and $\varepsilon_{\rm th}(T)$ are the thermal (kinetic)
contributions, it becomes clear
that the condensate contributions cancel one another on the r.h.s. of
eq.~(\ref{eq:ideal2-entropy-4}) and only the thermal contributions
with $\mu_I = m$ left in this equation.
%
\begin{figure}
\centering
\includegraphics[width=0.32\textwidth]{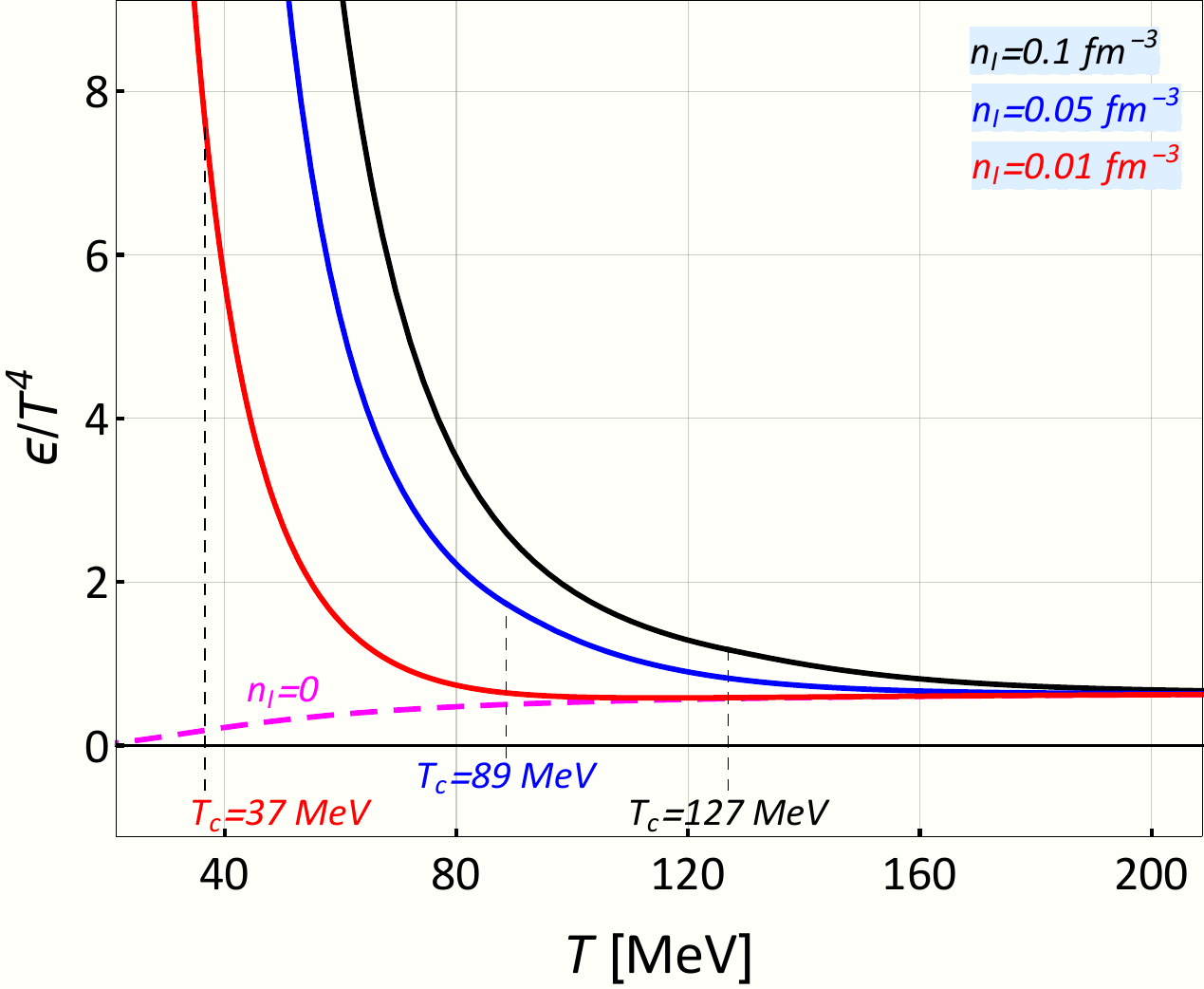}
\includegraphics[width=0.32\textwidth]{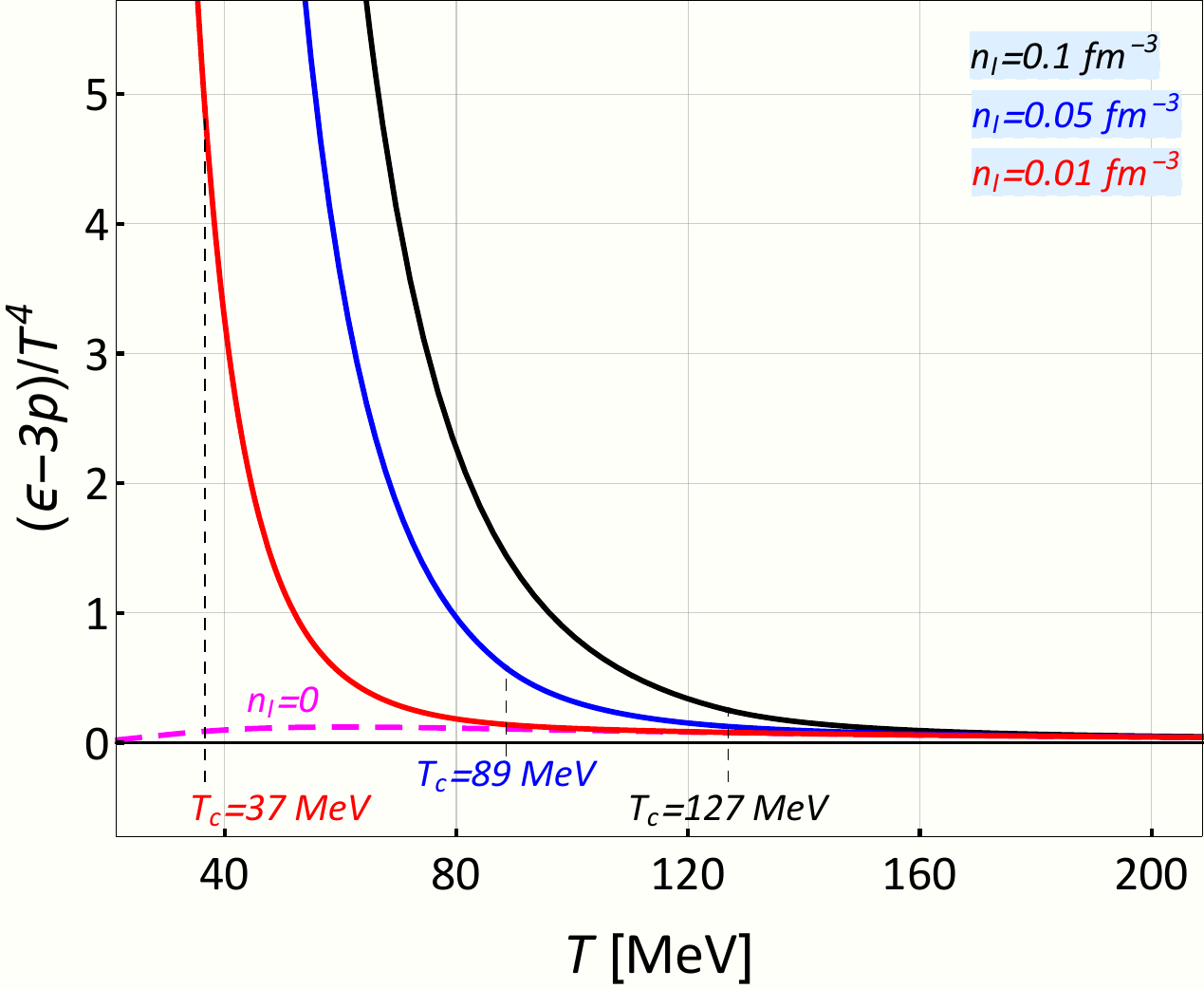}
\includegraphics[width=0.335\textwidth]{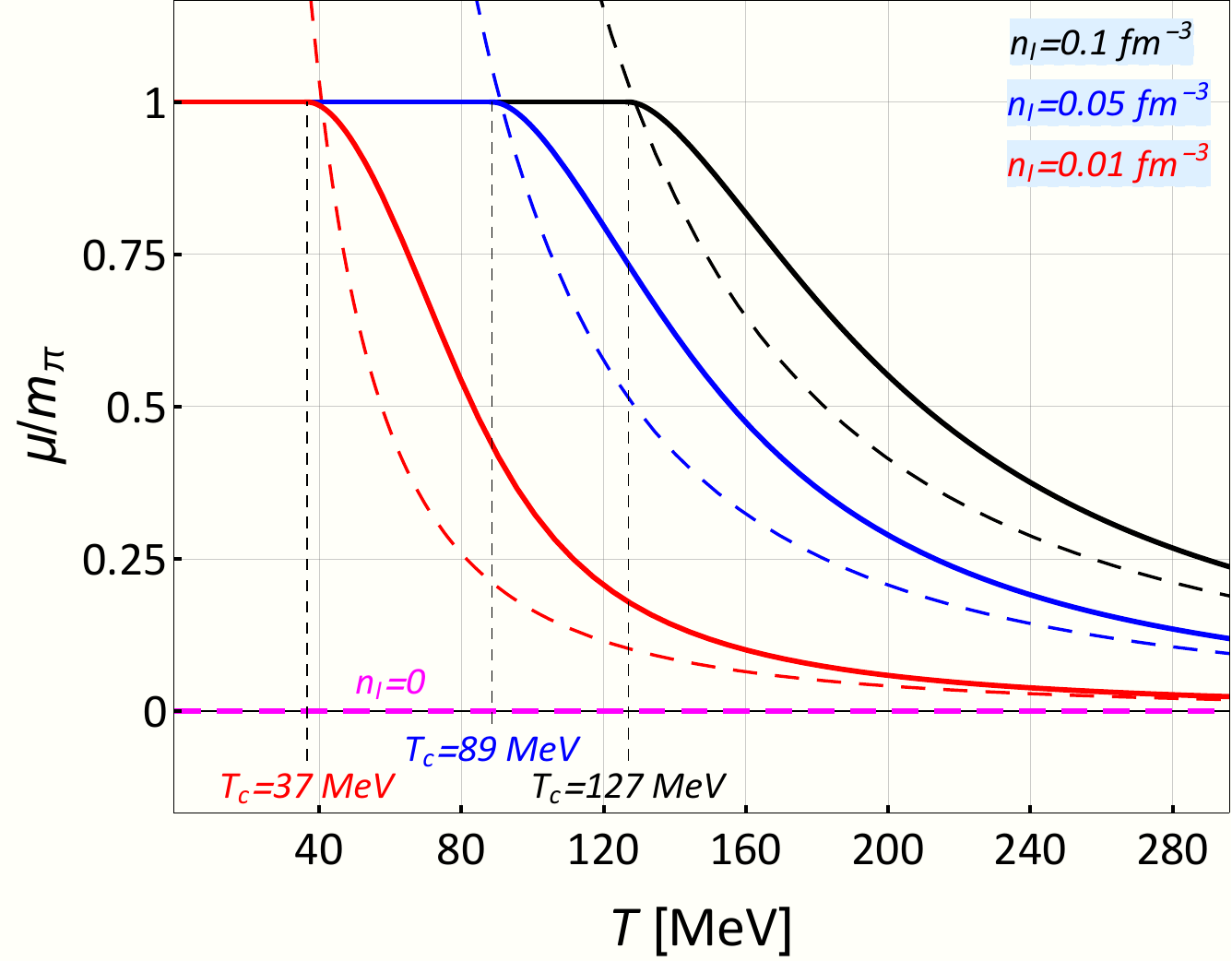}
\caption{An ideal charged bosonic gas of particles and antiparticles ($m = m_\pi$).
{\it Left panel:} Energy density vs. temperature
at $n_I = 0.1,\, 0.05,\, 0.01$~fm$^{-3}$, red, blue and black
lines respectively, and $n_I = 0$ as the dashed pink line.
The values of $T_{\rm c}$ shown correspond to a second-order phase transition.
{\it Central panel:} Temperature dependence of the trace anomaly
with the same parameter values and notation as in the left panel.
{\it Right panel:} Temperature dependence of the chemical potential.
The dashed lines correspond to the asymptotes ($T \to \infty$) of the chemical
potential $\mu_I=3n_I/T^2$ at different values of
$n_I = 0.01,\, 0.05,\, 0.1$~fm$^{-3}$.
The chemical potential $\mu_I = 0$, which is depicted as the dashed pink line,
corresponds to $n_I = 0$.
}
\label{fig:ideal-2comp-gas-phys-quantitie}
\end{figure}

The temperature dependencies of the particle density $n^{(-)}$ and
heat-capacity are shown in Fig.~\ref{fig:ideal-2comp-gas-particle-density},
in the left and right panels, respectively,
for the ideal $\pi^-$ - $\pi^+$ gas at the isospin densities
$n_I = 0.1,\, 0.05,\, 0.01$~fm$^{-3}$, and $n_I = 0$.
It is seen that the electrically neutral system with $n_I = 0$ does not
create condensate.
The temperature dependencies of the energy density and trace-anomaly
are shown in Fig.~\ref{fig:ideal-2comp-gas-phys-quantitie}, in the left and
right panels, respectively, for the same conditions as in
Fig.~\ref{fig:ideal-2comp-gas-particle-density}.
The break of the heat capacity at $T = T_{\rm c}$ shown in the right panel
in Fig.~\ref{fig:ideal-2comp-gas-particle-density}, together with a smoothness
behavior of the energy density at this temperature, that can be seen in the
left panel in Fig.~\ref{fig:ideal-2comp-gas-phys-quantitie},
signals about the phase transition of the second order at $T_{\rm c}$
in the system where $n_I \ne 0$.

At the same time, in Figs.~\ref{fig:ideal-2comp-gas-particle-density} and
\ref{fig:ideal-2comp-gas-phys-quantitie}, a discontinuity in the depicted
physical quantities at the point $n_I = 0$ is observed.
Indeed, it can be inferred from the temperature dependencies of the
functions $n^{(-)}(T,n_I)$, $c_{\rm v}(T,n_I)/T^3$, $\varepsilon(T,n_I)/T^4$
and $[\varepsilon(T,n_I) - 3p(T,n_I)]/T^4$, all of which exhibit a
discontinuity at $n_I \ne 0$.
We will discuss these features in the next Section \ref{sec:symmetry-breaking}.

\medskip

\noindent
{\it Asymptote of $\mu_I$ as $T \to \infty$.}

At high temperatures, one can derive an asymptotic expression for the chemical
potential $\mu_I(n_I,T)$.
To do this, we analyze the expression (\ref{eq:id-charge-density}) for the
isospin density $n_I$ at $T \to \infty$ (note that, at high temperatures,
there is no condensate contribution).
After performing an expansion similar to the one used to find the
Stefan-Boltzmann limit, and considering the asymptotic behavior
of the modified Bessel function $K_2$,
see eqs.~(\ref{eq:id2-sb-1a})-(\ref{eq:id2-sb-4}),  we arrive at
\begin{equation}
n_I \,\approx \, \frac{T^3}{\pi^2}\,\sum_{n = 1}^{\infty}\left(
e^{n\mu_I/T}-e^{-n\mu_I/T}\right)\frac{1}{n^3} \,
\approx \,  2\,\frac{\mu_I T^2}{\pi^2} \,\sum_{n = 1}^{\infty} \frac{1}{n^2} \,
=\, \frac{\mu_I T^2}{3} \,,
\label{eq:id-mu-1im2}
\end{equation}
where we take into account the value of the Riemann zeta function,
$\sum_{n = 1}^{\infty} 1/n^2 = \zeta(2) = \pi^2/6$.
Thus, we come at the asymptotic of the chemical potential
\begin{equation}
\lim_{T \to \infty} \, \mu_I(n_I,T) \,=\, \frac{3 n_I }{T^2} \,,
\label{eq:mu-1im3}
\end{equation}
which is consistent with the expression for the asymptotic of the
chemical potential $\mu_I$ obtained in Ref.~\cite{haber-1982}.
The asymptotes together with the exact temperature dependencies of the
chemical potential are shown in Fig.~\ref{fig:ideal-2comp-gas-phys-quantitie}
(right panel) for different values of $n_I = 0.01,\, 0.05,\, 0.1$~fm$^{-3}$.
The chemical potential corresponding to $n_I = 0$, i.e., $\mu_I = 0$,
is shown by the pink dashed line.

\section{Symmetry breaking due to the phase transition of the first order}
\label{sec:symmetry-breaking}

We are interested in the symmetry of the ground state $\Phi_0$ of the
particle-antiparticle system.
As stated in Section \ref{sec:lagrangian-ideal}, at zero temperature the ground
state is determined by eq.~(\ref{eq:eq-phi0-ideal}),
i.e. $\left(\partial_t^2 + m^2 \right) \Phi_0(t) = 0$, which has two solutions.
The first solution is trivial, $\Phi_0 = 0$; meanwhile, it defines a neutrally
charged two-component system.
As we have seen, this solution determines the absence of condensate in
the system of particles and antiparticles and automatically selects the
zero value of the isospin (charge) number, i.e. $n_I = 0$.
Consequently, there are no particles in the system at zero temperature;
a creation of particle-antiparticle pairs can be observed with an increase of
the temperature, see in Fig.~\ref{fig:ideal-2comp-gas-particle-density},
in the left panel, the dashed curve marked as $n_I = 0$.
The number of particles on the ground state, where particle momentum is zero,
is given by $N_{\rm gs} = 1/[\exp(m/T)-1]$.
In the thermodynamic limit as $V \to \infty$, the density of these particles
approaches zero, $n_{\rm gs} = N_{\rm gs}/V \to 0$.
Therefore, in a neutral particle-antiparticle system with $n_I = 0$ in the
thermodynamic limit, the particle and energy densities of the ground state
are zero because all created particles occupy thermal (kinetic) states.
From eqs.~(\ref{eq:cond-condensate-p}), (\ref{eq:cond-condensate-antip}) it
follows that in a neutrally charged two-component system, where $n_I = 0$ and,
therefore, $\mu_I = 0$, the formation of condensate, or the existing of a
macroscopic number of particle pairs in the ground state, is possible only
for a massless gas.
For a system of massive particles, this can be realized in the presence
of interaction between particle.
Indeed, as shown in \cite{mishustin-anchishkin-2019} for an interacting bosonic
particle-antiparticle system, in the case of strong attraction between particles,
there is a wide temperature interval where the quasiparticles have zero mass and
create condensate with both components.

Second solution of eq.~(\ref{eq:eq-phi0-ideal}) is given in
eq.~(\ref{eq:phi0-ideal}).
It represents a ``massive'' ground state of the system.
As we saw, in this case the energy density of the ground state of the system is
$\varepsilon_{\rm gs} = n_{\rm cond} m$.
Therefore, to create a ``massive'' ground state instead of the symmetric state,
$\Phi_0 = 0$, it is necessary to pump energy in the system, for instance
at $T = 0$.
It can be a pump, for example, of the negative particles to create a
negative charge in the bosonic system, or, what is the same, to create
$N_I \ne 0$.
The ``jump'' of energy $\Delta E = m N_I$ is nothing more as a latent heat which
was put into the system.
We argue that this transition from $\Phi_0 = 0$ to the ``massive'' ground state
of the system, where $N_I \ne 0$, should not be characterized as a
"spontaneous" symmetry breaking.
While this is indeed a transition from a symmetric ground state to a
symmetry-broken one, it is not spontaneous, as it results from the injection
of particles (energy) into the system by an external source that has influenced
our system of particles and antiparticles.

There are two possible conditions for the formation of condensate
$\mu_I = \pm m$.
First, consider $\mu_I = m$.
In this case, as we showed, the particles (let us say $\pi^-$ mesons)
create the condensate, which has a negative charge, hence
$n_{I\rm cond} = n_I^{(-)} > 0$.
Then, putting $\mu_I = m$ in eq.~(\ref{eq:id-charge-density}) we get that
the isospin density is positive, $n_I > 0$.
On the other hand, if $\mu_I = - m$, the antiparticles ($\pi^+$ mesons)
create the condensate, which has a positive charge and $n_{I\rm cond} < 0$.
Consequently, from eq.~(\ref{eq:id-charge-density}) follows that the isospin
density is negative, $n_I = n_I^{(+)} < 0$.
(We would like to figure out one point that can lead to misunderstanding:
in accordance with our choice of ``particles'' and ``antiparticles'',
the negative extra charge in the system corresponds to a positive isospin
density $n_I > 0$,
and vice versa, the positive extra charge corresponds to a negative isospin
density $n_I < 0$.)

One can write the density of the free energy
(\ref{eq:charge-cond-legendre-transform-2c}) in the condensate phase
for the two conditions of the condensate creation:
1) $n_I > 0$ that determines $\mu_I = + m$, and
2) $n_I < 0$ that determines $\mu_I = - m$.
We get
\begin{eqnarray}
\label{eq:cond-dens-free-energy-neg}
n_I > 0,  \quad
\Phi(T,n_I)  &=&   m  n_I + T \int \frac{d^3k}{(2\pi)^3}
\left[ \ln{ \left( 1- e^{\frac{- \omega_k + m}{T} } \right) }
+ \ln{ \left( 1- e^{\frac{- \omega_k - m}{T} } \right) } \right]  ,
\\
n_I < 0,  \quad
\Phi(T,n_I)  &=&  (- m)n_I + T \int \frac{d^3k}{(2\pi)^3}
\left[ \ln{ \left( 1- e^{\frac{- \omega_k - m}{T} } \right) }
+ \ln{ \left( 1- e^{\frac{- \omega_k + m}{T} } \right) } \right]  ,
\label{eq:cond-dens-free-energy-pos}
\\
n_I = 0,  \quad
\Phi(T,n_I)  &=&  2 T \int \frac{d^3k}{(2\pi)^3} \,
\ln{ \left( 1- e^{- \omega_k/T } \right) } \,,
\label{eq:cond-dens-free-energy-neutral}
\end{eqnarray}
where in the third line we added the case of the neutral system $n_I = 0$,
which is realized at $\mu_I = 0$.
Here to complete eq.~(\ref{eq:cond-dens-free-energy-neg}), where $\mu_I = m$,
we take into account in eq.~(\ref{eq:charge-cond-legendre-transform-2c}) that:
$n_{I\rm cond} = n_{\rm cond}^{(-)} - n_{\rm cond}^{(+)}$ with
$n_{\rm cond}^{(+)} = 0$, and it is also $n_{\rm cond} = n_{\rm cond}^{(-)}$.
Hence, we wrote in (\ref{eq:cond-dens-free-energy-neg})
\[ (\varepsilon_0 n_{\rm cond} - \mu_I n_{I\rm cond}) + \mu_I n_I
= n_{\rm cond}^{(-)} (m - \mu_I) + m n_I = m n_I \,.
\]
To complete eq.~(\ref{eq:cond-dens-free-energy-pos}), where $\mu_I = - m$,
we take into account in eq.~(\ref{eq:charge-cond-legendre-transform-2c}) that:
$n_{I\rm cond} = - n_{\rm cond}^{(+)}$ because
$n_{\rm cond}^{(-)} = 0$, and in addition $n_{\rm cond} = n_{\rm cond}^{(+)}$.
Hence, we wrote in (\ref{eq:cond-dens-free-energy-pos})
\[ (\varepsilon_0 n_{\rm cond} - \mu_I n_{I\rm cond}) + \mu_I n_I
= n_{\rm cond}^{(+)} (m + \mu_I) - m n_I = - m n_I \,.
\]

\begin{figure}
\centering
\includegraphics[width=0.32\textwidth]{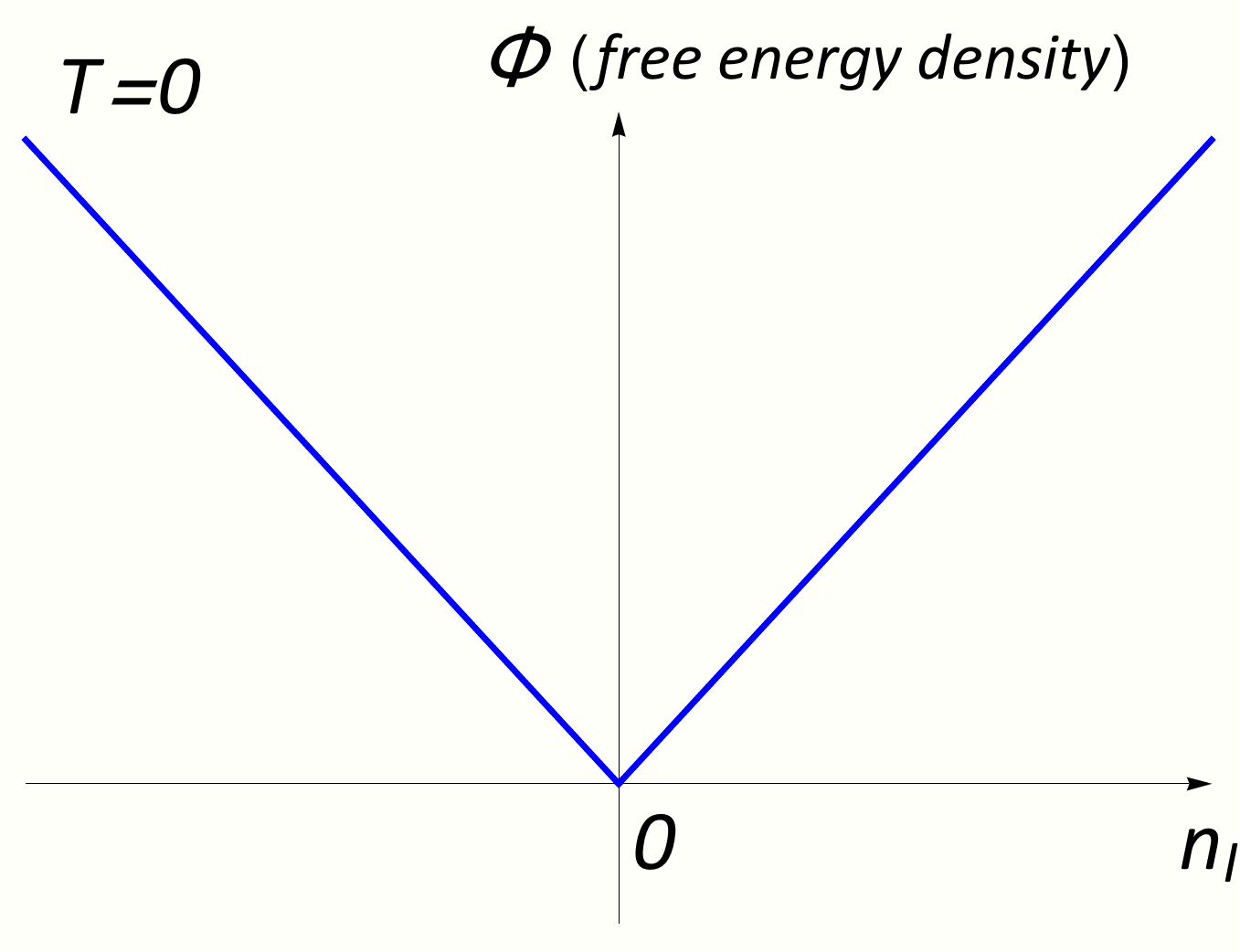}
\includegraphics[width=0.32\textwidth]{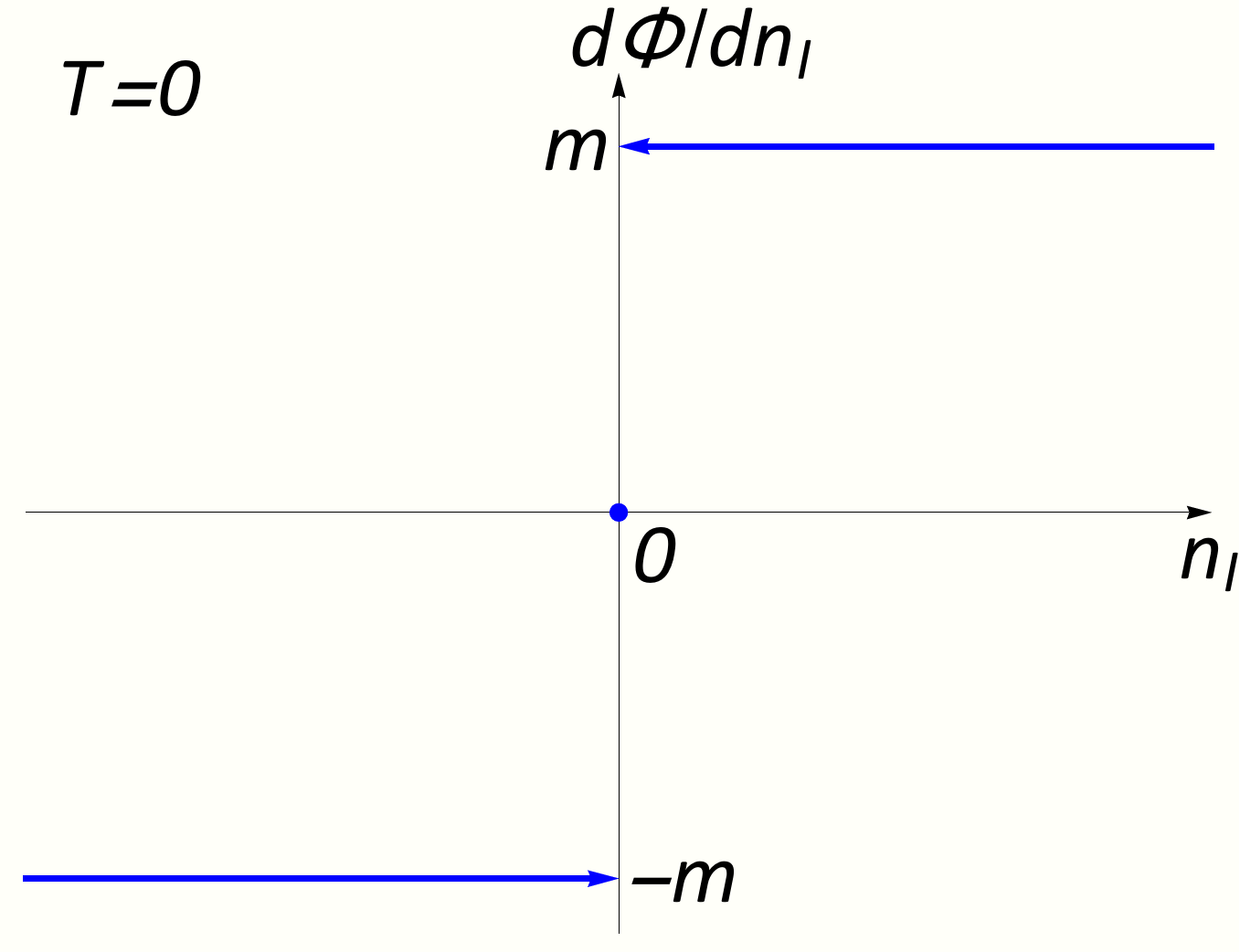}
\includegraphics[width=0.32\textwidth]{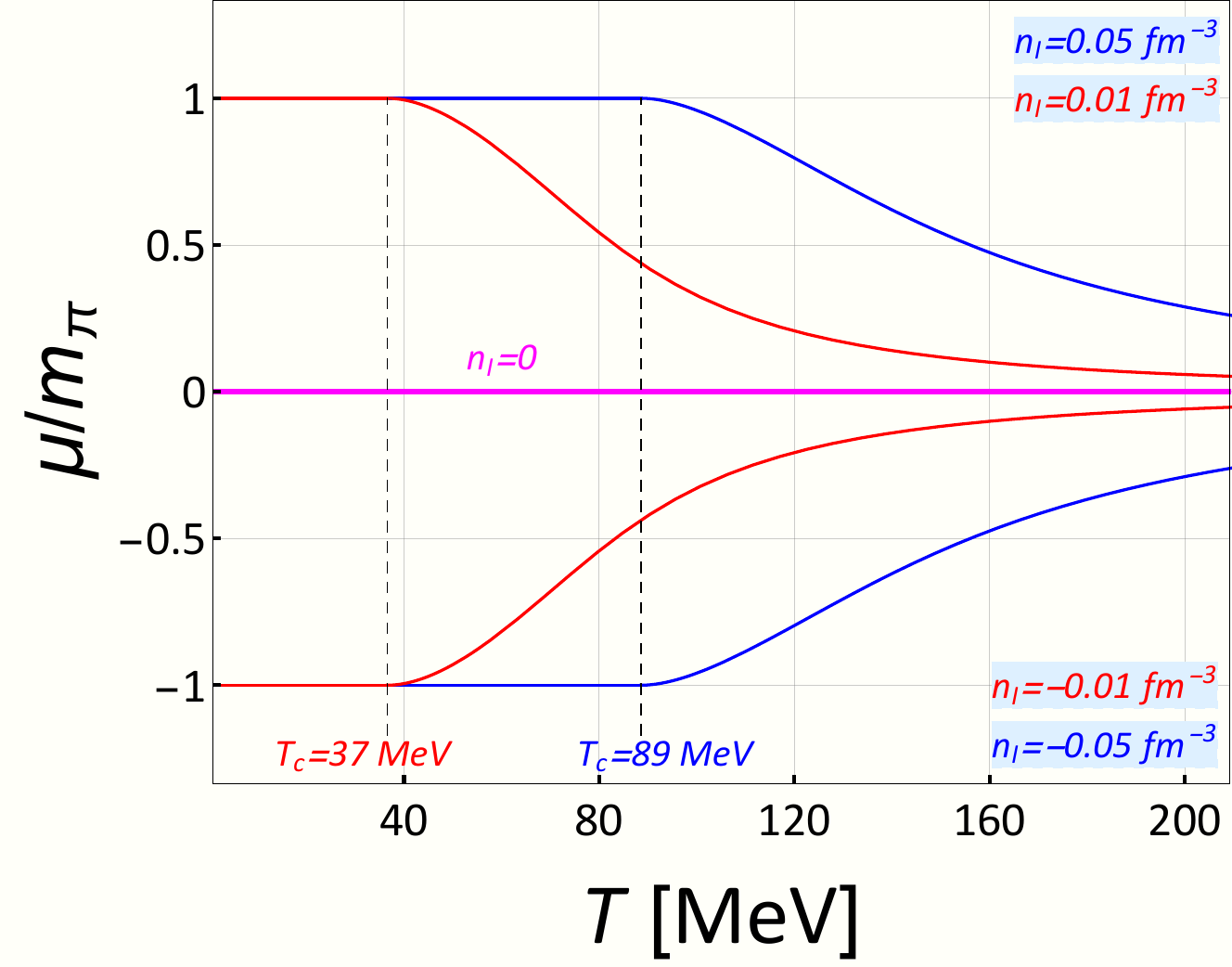}
\caption{An ideal charged bosonic gas of particles and antiparticles.
{\it Left panel:} Free energy density vs. isospin density at $T = 0$.
{\it Central panel:} The derivative of the free energy density vs. the
isospin density at $T = 0$.
{\it Right panel:} Temperature dependence of the chemical potential
at $n_I = 0.1,\, 0.05,\, 0.01$~fm$^{-3}$, red, blue and black
lines respectively, and $n_I = 0$ as the dashed pink line.
The values of $T_{\rm c}$ shown correspond to a second-order phase transition.
 }
\label{fig:free-energy}
\end{figure}

It is necessary to point out that the derivative of the density of the free
energy (\ref{eq:cond-dens-free-energy}) with respect to isospin density $n_I$
has a discontinuity in the point $n_I = 0$.
Indeed, at zero temperature, $T = 0$, it is true that
(see in Fig.~\ref{fig:free-energy} the left and central panels):
\begin{eqnarray}
\label{eq:id-derivative-positive-ni}
& n_I > 0 \quad  \rightarrow  \quad
 \frac{\dis \partial \Phi(T,n_I)}{\dis \partial n_I} = \mu_I = m \,, \quad
& n_{\rm cond} = \left(\psi_0^{(-)}\right)^*\psi_0^{(-)}\,,
\\
& n_I < 0 \quad  \rightarrow  \quad
 \frac{\dis \partial \Phi(T,n_I)}{\dis \partial n_I} = \mu_I = - m \,, \quad
& n_{\rm cond} = \left(\psi_0^{(+)}\right)^*\psi_0^{(+)} \,,
\label{eq:id-derivative-negative-ni}
\\
& n_I = 0 \quad  \rightarrow  \quad
 \frac{\dis \partial \Phi(T,n_I)}{\dis \partial n_I} = \mu_I = 0 \,, \quad
& n_{\rm cond} = 0 \,,
\label{eq:id-derivative-zero-ni}
\end{eqnarray}
where to give the value of  $n_{\rm cond}$ we use eq.~(\ref{eq:cond-contrib-2})
and the fact that only one component of the particle-antiparticle system can
develop a condensate.
Based on the behavior of the function $\Phi(n_I)$ at zero
temperature, as shown in Fig.~\ref{fig:free-energy} in the left
panel, the derivative of this function at the point $n_I = 0$ is indefinite.
However, from eq.~(\ref{eq:id-charge-density}) in the point $n_I = 0$ we obtain
that $\mu_I = 0$ and because $\mu_I = \partial \Phi(T,n_I)/\partial n_I$, we can
naturally set in the point $n_I = 0$ the value of the derivative as:
$\partial \Phi(n_I)/\partial n_I = \mu_I = 0$.
Exactly this statement is formulated in eq.~(\ref{eq:id-derivative-zero-ni})
and it is indicated as a blue point in the beginning $n_I = 0$
in Fig.~\ref{fig:free-energy} in the central panel.
%

So, we fix a discontinuity (gap) of the function $\mu_I(T,n_I)$
in the point $n_I = 0$, which takes place in the temperature interval
$T < T_{\rm c}$.
In accordance with the Ehrenfest classification \cite{ehrenfest-1933}
this discontinuity should be treated as phase transition of the first order.
In addition, the gap of the function $\frac{\partial \Phi(T,n_I)}{\partial n_I}$
means that all thermodynamic quantities that are calculated using the free energy
$\Phi$
have a discontinuity in the point $n_I = 0$ for temperatures $T < T_{\rm c}$.
We see this peculiarities in Fig.~\ref{fig:ideal-2comp-gas-particle-density} and
Fig.~\ref{fig:ideal-2comp-gas-phys-quantitie}, where the results of calculation
of the dependence with respect to temperature of the particle density $n^{(-)}$,
energy density, heat capacity and trace anomaly are shown.
This specific gap-like behavior arises from the analytical expression for
the free energy given in eq.~(\ref{eq:cond-dens-free-energy}).
Indeed, if $n_I \ne 0$, the right-hand side of expression
(\ref{eq:cond-dens-free-energy}) has an additive term proportional to
$n_{\rm cond}(T)$, where $n_{\rm cond}(T=0) = n_I$.
Thus, thermodynamic quantities that involve derivatives of the free energy
with respect to temperature $T$ acquire an additional term proportional to
$\propto  \partial n_{\rm cond}(T)/\partial T$.
In a neutral system, i.e., $n_I = 0$, the term proportional to
$n_{\rm cond}(T)$ is excluded from the free energy, and there
is no extra contribution from the condensate energy.

If we consider the bosonic system at $n_I > 0$ that means $\mu_I = m$,
from eqs.~(\ref{eq:id-derivative-positive-ni}) and (\ref{eq:id-derivative-zero-ni})
we fix the gap of the derivative of free energy, what
is nothing but a phase transition of the first order in the point $n_I = 0$.
The finite isospin density, $n_I \ne 0$, which is associated with a
nontrivial ground state, results in the creation of condensate at $T = 0$.
At the same time, when the ground state of the system at $T = 0$ is trivial,
$\Phi_0 = 0$, hence the isospin (charge) density is zero, $n_I = 0$,
there is no condensate phase in an ideal particle-antiparticle boson gas.
This means that the formation of a condensate or the injection of charged
particles into the system at $T = 0$, which is nothing more than a phase
transition of the first order at this point, leads to a breaking of
the symmetry of the ground state:
\begin{equation}
\Phi_0 = 0 \quad \rightarrow \quad \Phi_0
=  \frac{e^{-imt}}{\sqrt{2m}} \, \psi_0^{(-)} \,.
\label{eq:gs-pi-minus}
\end{equation}
Here on the r.h.s. of this equation, we present the solution for the ground
state (\ref{eq:phi0-ideal}), where we keep only one component that has a
dominant charge and creates a condensate with the scalar density
%
\begin{equation}
\Phi_0^* \Phi_0 \,=\, \frac{1}{2m} \, n_I
\qquad  {\rm with}  \qquad
n_I \,=\, \left(\psi_0^{(-)}\right)^* \psi_0^{(-)} \,.
\label{eq:gs-density}
\end{equation}
This relation connects the scalar density of condensate with the isospin
density $n_I$ at $T = 0$.
At finite temperatures, the isospin density $n_I$ also incorporates the
thermal fraction of the isospin density
$n_{I {\rm th}} = n_{\rm th}^{(-)} - n_{\rm th}^{(+)}$.
From relation (\ref{eq:gs-density}), it follows that
the symmetric ground state $\Phi_0 = 0$, or an absence of the
condensate, corresponds to the absence of the extra charge in the system,
$n_I = 0$, which aligns with eq.~(\ref{eq:id-derivative-zero-ni}).

\begin{figure}
\centering
\includegraphics[width=0.49\textwidth]{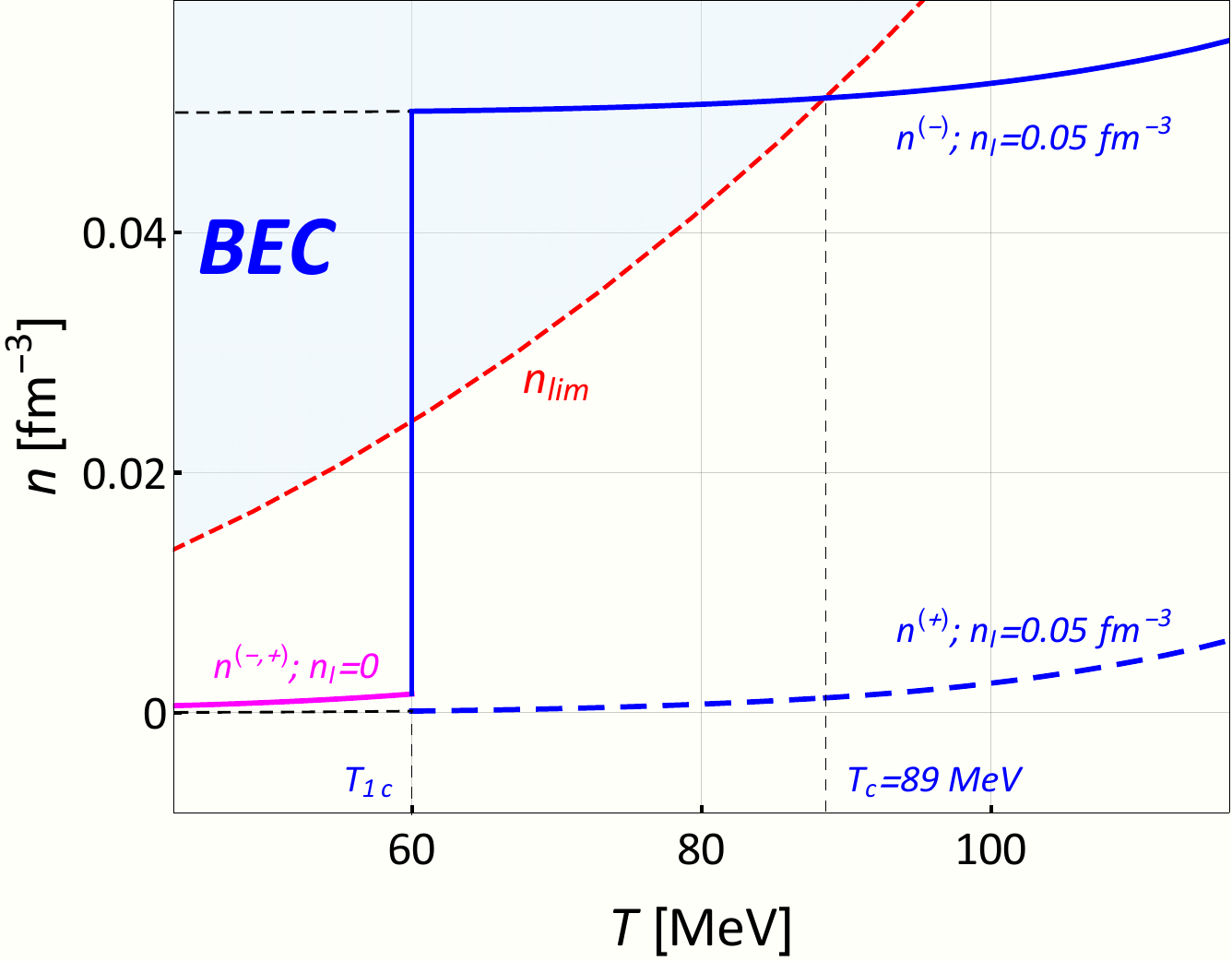}
\includegraphics[width=0.49\textwidth]{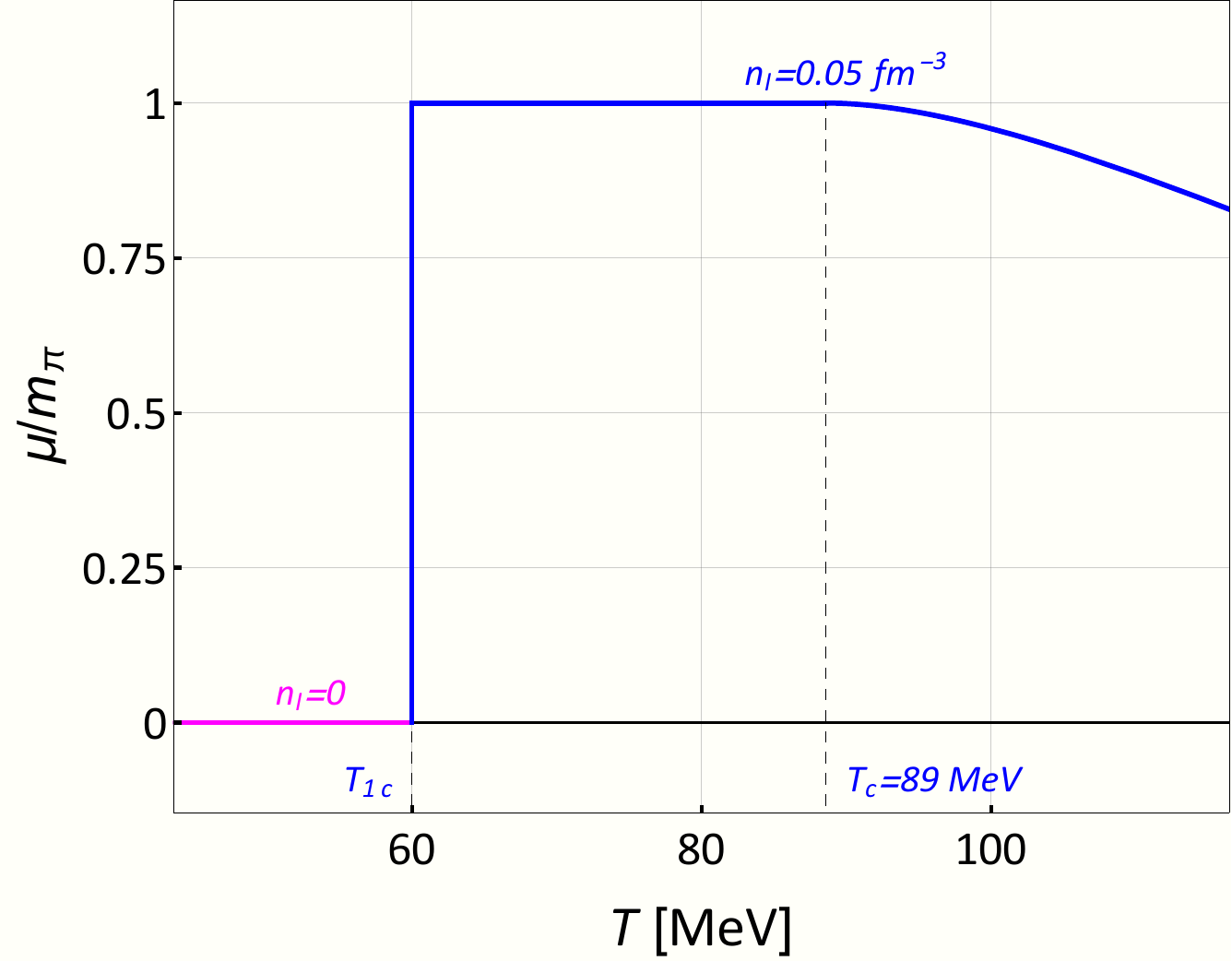}
\caption{An ideal charged bosonic gas of particles and antiparticles.
{\it Left panel:} Particle-number density vs. temperature.
{\it Right panel:} Temperature dependence of the chemical potential
at $n_I = 0.05$~fm$^{-3}$, blue line.
The temperature of $T_{1\rm c}$ corresponds to a first-order phase transition.
The temperature of $T_{\rm c}$ corresponds to a second-order phase transition.
 }
\label{fig:first-order-phase-tr}
\end{figure}

We can go further and consider the symmetry breaking at finite temperatures.
Let us begin heating a charge-neutral bosonic system, which consists of
particles and antiparticles, from zero temperature.
The ground state $\Phi_0 = 0$ is not broken, indicating that there is no
condensate in the system, the chemical potential is zero ($\mu_I = 0$),
and the equilibrium thermodynamic states of the system are described by the
free energy given in eq.~(\ref{eq:cond-dens-free-energy-neutral}).
At a temperature $T_{1 \rm c}$, we start pumping into the system, for
instance, negative particles, which create an additional charge with
a density of $n_I$.
To achieve thermal overpopulation with negatively charged particles, which
is a condition for the formation of condensate, the additional isospin density
$n_I$ must be greater than
$n_I > n_{\rm lim}(T_{1 \rm c}) - n_{\rm th}^{(-)}(T_{1 \rm c})$,
where $n_{\rm lim}(T)$ represents the density of negative particles on the
critical curve, and $n_{\rm th}^{(-)}(T_{1 \rm c})$ denotes the density of
negative particles before pumping.
The pumping of additional negative particles signifies an external
influence on the bosonic system, similar to heating water by an external
hot system, and if the heat transferred is sufficient, boiling occurs,
i.e. a phase transition.
The stages of increasing the particle density are depicted in
Fig.~\ref{fig:first-order-phase-tr} in the left panel.
The process of pumping additional particles into the system leads to
a ``jump'' of energy density $\Delta \varepsilon(T_{1 \rm c}) = m n_{\rm cond}
+ \Delta \varepsilon_{\rm th}^{(-)}$.
After the system relaxes to thermodynamic equilibrium, its states are
described by the free energy given in eq.~(\ref{eq:cond-dens-free-energy-neg}).
It is seen that the system ``earns'' a finite chemical potential $\mu_I \ne 0$,
which is shown in Fig.~\ref{fig:first-order-phase-tr} in the right panel.
So, by introducing additional energy (particles) into the system
at $T = T_{1 \rm c}$, we create a ``condensate'' ground state instead of the
symmetric state $\Phi_0 = 0$ that characterized the system before the
external influence.
So, we can say that this transition from $\Phi_0 = 0$ to the ``condensate''
ground state with a broken symmetry is a first-order phase transition.
Indeed, the transition is characterized by a) a ``jump'' in the energy density,
which is simply latent heat introduced into the system, and
2) a break in the derivative of the free energy with respect to $n_I$ at the
temperature $T_{1 \rm c}$, see Fig.~\ref{fig:first-order-phase-tr}, right panel.
The transition occurs due to the injection of particles (energy) into the
system by an external source, which also impacts our system of particles
and antiparticles by increasing the pressure (we consider the system at
constant volume $V$).
A reversible first-order phase transition, accompanied by the release of
latent heat, can occur when the temperature decreases along with
decreasing pressure, for example, at $T_{1 \rm c}$.

\section{Conclusions and discussion}
\label{sec:conclusions}

In this study, we presented a thermodynamic and field-theoretic
analysis of relativistic Bose-Einstein condensation (BEC) in an ideal gas
of bosons and antibosons with conserved charge (isospin).
Our approach, based on a scalar field model and extended canonical
ensemble, allows a precise treatment of the chemical potential as a
thermodynamic function of temperature and charge density.
Within this framework, we confirm the central conclusions of
Kapusta~\cite{kapusta-1981} and Haber and Weldon~\cite{haber-1982},
who interpret BEC as a manifestation of finite-temperature symmetry
breaking in scalar field theory.
However, in contrast to their spontaneous symmetry-breaking scenario, we find
that in the ideal gas, the condensate emerges only under the external injection
into the system of a finite charge.
Consequently, the symmetry breaking is not spontaneous but induced.

Our results are also consistent with those of Begun and
Gorenstein~\cite{begun-2007,begun-2008}, who emphasize the role of conserved
charges and ensemble choice in the context of relativistic BEC.
However, our formalism provides a more explicit and general procedure for
transitioning from the grand canonical to the Canonical Ensemble,
thereby clarifying the thermodynamic constraints that arise in the
condensate phase.
A development of the extended CE framework to include fluctuation observables
and finite-volume corrections is a promising direction for future research.

Our study was conducted within the extended Canonical Ensemble framework,
where the chemical potential $\mu_I$ is treated as a thermodynamic function
of conserved charge and temperature.
We arrived at this formalism by first examining the system in the grand
Canonical Ensemble and then performing a Legendre transformation.
We showed that the Grand Canonical Ensemble becomes inadequate in the
condensate phase without this modification, thus supporting the conclusions
of Refs.~\cite{universe-2023,ujp-2024}, which considered bosonic systems
both with and without interactions.
This limitation arises because, in the condensate phase, the chemical potential
is no longer a free parameter but is fixed by the condition for condensate
formation, specifically, $\mu_I = m$ for the ideal gas of particles and
antiparticles considered in this work.

At zero temperature, the particle-antiparticle bosonic system admits two
possible ground states: the first is an "empty" ground state with $\Phi_0 = 0$,
and the second is a "massive" ground state with energy density
$\varepsilon_0 = n_I m$ and finite isospin $N_I = V n_I \ne 0$.
We argue that the transition from the neutral ground state $\Phi_0 = 0$ to
the charged ground state cannot be classified as spontaneous.
While it involves symmetry breaking, the transition is induced by the injection
of particles (or energy) into the system from an external source.

We calculated the discontinuity in the first derivative of the free energy,
$\partial \Phi(T,n_I)/\partial n_I$, at the point $n_I = 0$, which
corresponds to the symmetric ground state $\Phi_0 = 0$.
This reveals a discontinuity in the chemical potential as a function of $n_I$
at this point and, correspondingly, discontinuities in other thermodynamic
quantities.
Specifically, the energy density exhibits a jump from $\varepsilon = 0$ to
$\varepsilon = m n_I$ at fixed temperature $T = 0$, since the injected
negative particles occupy the zero-momentum state.
According to the Ehrenfest classification~\cite{ehrenfest-1933}, such a
discontinuity qualifies as a first-order phase transition.

Thus, we demonstrate that the formation of a condensate at $T = 0$, along with
the emergence of finite isospin $N_I = N_{\rm cond} \ne 0$, or equivalently,
the transition between the neutral and charged ground states, constitutes
a first-order phase transition in the particle-antiparticle bosonic system.
This first-order behavior at $n_I = 0$, linked to the discontinuity in $\mu_I$,
reveals a novel thermodynamic feature of the relativistic ideal Bose gas - one
that has not been emphasized in previous literature.
This insight highlights the distinctive behavior of particle-antiparticle
systems and may have implications for interpreting condensate formation in
systems with charge or isospin asymmetry.

We note that spontaneous symmetry breaking can still occur in interacting
systems when the effective potential develops a nontrivial (e.g., negative)
minimum.
Such behavior is well known, for instance, in models with a Mexican hat
potential at zero temperature.

Finally, we emphasize that in a multicomponent bosonic system, a Bose-Einstein
condensate forms only in the component with the dominant particle density.
This observation is relevant to pion gases created in high-energy heavy-ion
collisions, where an excess of negatively charged pions is typically observed:
$n_I = n_\pi^{(-)} - n_\pi^{(+)} > 0$.
This implies that only $\pi^{-}$ mesons undergo condensation into the ground
state at zero momentum $\bs k = 0$, and hence, the condensate carries negative
electric charge.
Meanwhile, $\pi^{+}$ mesons remain entirely in the thermal phase across all
temperatures.

This scenario closely resembles superfluidity: only the condensate contributes
to collective flow.
The $\pi^{+}$ mesons and the thermal fraction of the $\pi^{-}$ mesons form
the "normal fluid", while the superfluid component is made up of condensed
$\pi^{-}$ mesons, collectively carrying a negative charge.
From an experimental perspective, it is therefore meaningful to analyze the
total charge of pions participating in collective flow.
If this total charge is negative, it signals the formation of a Bose-Einstein
condensate of $\pi^{-}$ mesons during the fireball evolution stage of a
relativistic nucleus-nucleus or particle-nucleus collision.
Conversely, a positive total charge would imply the condensation of $\pi^{+}$
mesons under similar conditions.

\section*{Acknowledgements}
We thank V.~Gusynin for very useful and instructive discussions.
Our thanks to Yu.~Slyusarenko, M.~Shpot, and V.~Skalozub for the valuable
remarks.
We also thank D.~Zhuravel and V.~Karpenko for cooperation.
The work of D.~A. and V.~Gn. is supported by the Simons Foundation and
by the Program "The structure and dynamics of statistical and
quantum-field systems" of the Department of Physics and Astronomy of
the NAS of Ukraine.
This work is supported also by Department of
target training of Taras Shevchenko Kyiv National University and the
NAS of Ukraine, grant No. 6$\Phi$-2024.

\section*{Appendix}
\label{sec:appendix}
\numberwithin{equation}{subsection}

\subsection{Exact conservation of particles in a single-component ideal
bosonic gas in the thermal phase}
\label{sec:legendre-transform-single-component}

It is assumed that the system is subject to the law of conservation of
the number of particles.
To ensure exact conservation of the number of particles $N$, we first consider
the system in the Grand Canonical Ensemble (GCE) and then move to the Canonical
Ensemble through the Legendre transformation.
This method provides a formally correct scheme for incorporating the chemical
potential at the microscopic level into the description of the system within
the Canonical Ensemble framework.
Indeed, we begin from the partition function in GCE that reads
\begin{equation}
Z(T,\mu,V) \,=\, {\rm Tr} \exp{\left[-(H - \mu \hat N)/T \right]} \,,
\label{eq:part-func-id-gas}
\end{equation}
where $H = \sum_{\bs k} \omega_{\bs k}\left(a^+_{\bs k} a_{\bs k}
+ b^+_{\bs k} b_{\bs k}\right)$ with $\omega_k = \sqrt{m^2 + \bs k^2}$, and
$\hat N = \sum_{\bs k} \left(a^+_{\bs k} a_{\bs k} + b^+_{\bs k} b_{\bs k}\right)$.

For the sake of simplicity we consider a single-component ideal system of bosons
for the temperature interval $T > T_{\rm c}$, where $T_{\rm c}$
is the critical temperature at which condensate formation begins.
In the GCE we have the following relation for the number of particles in the
system
\begin{equation}
N \,=\, -\, \left[ \frac{\partial \Omega(T,\mu,V)}{\partial \mu} \right]_{T,V}\,,
\label{eq:legendre-transform-1}
\end{equation}
where $\Omega$ is the thermodynamic potential that in explicit form reads
\begin{eqnarray}
\Omega(T,\mu,V) \,=\, T V \int \frac{d^3k}{(2\pi)^3} \,
\ln{ \left[ 1- e^{- \big(\omega_k - \mu \big)/T } \right] } \,.
\label{eq:omega-therm-single-particle}
\end{eqnarray}
%
We are going to make the Legendre transformation.
We solve equation (\ref{eq:legendre-transform-1}) with respect to $\mu$ and
get a function $\mu = \mu(T,N/V)$.
Next we define the function $F$
\begin{equation}
F(T,N,V) \,=\, \left[\Omega \,
-\, \mu \, \frac{\partial \Omega}{\partial \mu}\right]_{T,V} \,,
\label{eq:legendre-transform-2}
\end{equation}
where it is implied that the chemical potential is a function of
variables $(T,\frac NV)$.
In explicit form this equation reads
\begin{equation}
F(T,N,V) \,=\, -T \, \ln{\left\{{\rm Tr} \,
\left[e^{\dis -(H - \mu(T,n) \hat N)/T}\right]\right\}} \,+\, \mu(T,n) \, N \,,
\label{eq:legendre-transform-2.1}
\end{equation}
where $n = N/V$.
Equation (\ref{eq:legendre-transform-2}) can be rewritten as (we exploit that
in a homogeneous system $\Omega(T,\mu,V) = - p(T,\mu) V$)
\begin{equation}
F(T,N,V) \,=\, - p\big(T,\mu(T,N/V)\big)\, V \,+\, \mu(T,N/V) \, N  \,,
\label{eq:legendre-transform-2a}
\end{equation}
and we immediately recognize the free energy that is a function of
the free variables $(T,N,V)$, whereas the chemical potential is now
a thermodynamic function, $\mu(T,N/V)$.
For example, from eq.~(\ref{eq:legendre-transform-2}) with a use of
eq.~(\ref{eq:legendre-transform-1}) follows
\begin{equation}
\mu \,=\, \frac{\partial F(T,N,V)}{\partial N} \,.
\label{eq:legendre-transform-5}
\end{equation}
In an explicit form eq.~(\ref{eq:legendre-transform-2a}) looks like
\begin{equation}
\Phi(T,n) \,=\, T \int \frac{d^3k}{(2\pi)^3}
\ln{ \left[ 1- e^{- \big(\omega_k - \mu \big)/T } \right] } \,+\, \mu \, n \,,
\label{eq:legendre-transform-2c}
\end{equation}
where $\Phi(T,n) = F(T,N,V)/V$ is the density of the free energy.
The chemical potential as a function of the conserved particle-number
density $n$ and temperature $T$,
is obtained as solution of eq.~(\ref{eq:legendre-transform-1}), which in an
explicit form reads
\begin{equation}
n  \,=\,  \int \frac{d^3k}{(2\pi)^3}\, f_{\rm BE}\big(\omega_k,\mu\big) \,.
\label{eq:thermal-particle-density-1}
\end{equation}
A knowledge of the chemical potential as the function of temperature and
particle-number density opens the way to calculate other thermodynamic
quantities using the extended statistical operator
$\rho(T,n) = \exp{\left[-\left( H - \mu(T,n) \hat N \right)/T\right]}$.
In particular, this approach gives us a possibility to consider the problems
where the particle-number density is kept constant, $n =$~const.

\bigskip

One can argue that within the Canonical Ensemble, there is a well-known
thermodynamic relation $dF = - s dT - p dV + \mu dN$ for describing a system
with particle-number variations.
Therefore, the chemical potential can be defined using equation
eq.~(\ref{eq:legendre-transform-5}).
Then, there seems to be no need for the scheme proposed above.
However, it should be noted that eq.~(\ref{eq:legendre-transform-5}) defines
the chemical potential in the thermodynamic approach; it does not provide
a recipe for using the chemical potential in statistical microscopic
calculations, nor a method for incorporating the chemical potential into
the system's description when the number of particles is kept constant,
$N =$~const, across different temperatures.

At the same time, the method proposed in the present paper provides a formally
correct scheme for incorporating the chemical potential into the description
of the system {\it at the microscopic level} within the Canonical Ensemble
framework.
Indeed, the thermodynamic properties of the bosonic system are analyzed
on base of the quantum statistical averaging giving in
eq.~(\ref{eq:legendre-transform-2.1}) that can be named as the Extended
Canonical Ensemble (ECE).
It is seen that in the ECE the mean values are obtained now with the use of
the statistical operator
%
\begin{equation}
\rho(T,n) \,=\, e^{ - \left(H - \mu(T,n) \hat N\right)/T} \,,
\label{eq:ece-stat-oper}
\end{equation}
where the temperature $T$ and the particle-number density $n$ are given
quantities.
Therefore, the partition function, calculated using the extended statistical
operator (\ref{eq:ece-stat-oper}), along with an understanding of the chemical
potential as a function of temperature and particle-number density, opens
the way for the microscopic calculation of all thermodynamic quantities.
In particular, this approach allows us to consider problems where the
particle-number density exactly remains constant, $n =$~const, at different
temperatures.

\subsection{\bf Another approach to ensure the conservation of the isospin
(charge) number in the particle-antiparticle boson system in the Canonical
Ensemble framework}
\label{sec:kronecker-isospin-id-gas}
In the framework of the Canonical Ensemble the mean occupation numbers for
particles $n_k^{(-)}$ and antiparticles $n_k^{(+)}$ read
\begin{eqnarray}
\label{eq:mean-number-minos0}
n_k^{(-)} \, \equiv \, \langle a^+_k a_k \rangle &=&
\frac1Z \, {\rm Tr}\left[ e^{- \beta H}\, a^+_k a_k \, \right] \,,
\\
n_k^{(+)} \, \equiv \, \langle b^+_k b_k \rangle &=&
\frac1Z \, {\rm Tr}\left[ e^{- \beta H}\, b^+_k b_k \, \right] \,.
\label{eq:mean-number-plus0}
\end{eqnarray}
We are going to consider the system in which the isospin
$N_I = N^{(-)} - N^{(+)}$ is kept constant, where
$N^{(-)} = \sum_{\bs k} N^{(-)}_{\bs k}$,
$N^{(+)} = \sum_{\bs k} N^{(+)}_{\bs k}$.
To ensure an exact conservation of the isospin (charge) number one can introduce
the Kronecker symbol $\Delta\left(N^{(-)} - N^{(+)} - N_I\right)$ into the
formulae (\ref{eq:mean-number-minos0}) and (\ref{eq:mean-number-plus0})
for the mean values in the following way
\begin{eqnarray}
\label{eq:mean-number-minos2}
n_k^{(-)} &=& \frac1Z \, {\rm Tr}\left[ e^{- \beta H}\,
a^+_k a_k \, \Delta\left(N^{(-)} - N^{(+)} - N_I\right)\right] \,,
\\
n_k^{(+)} &=& \frac1Z \, {\rm Tr}\left[ e^{- \beta H}\,
b^+_k b_k \, \Delta\left(N^{(-)} - N^{(+)} - N_I\right) \right] \,.
\label{eq:mean-number-plus2}
\end{eqnarray}
Then, it is reasonable to use a representation of the Kronecker function:
\begin{equation}
\Delta\left(N^{(-)} - N^{(+)} - N_I\right) = \frac{\beta}{2\pi i}\int_0^{2\pi i}
d\alpha \, e^{ \beta \alpha \left(N^{(-)} - N^{(+)} - N_I\right)} \,,
\label{eq:kroneker-repr}
\end{equation}
where $\beta =$~const and in our case $\beta = 1/T$.

\bigskip

\noindent
{\it Particles and antiparticles are in a thermal phase  ($T > T_{\rm c}$).}
\\
We are going to consider the ideal boson gas with the Hamiltonian
$H = \sum_{\bs k} \omega_{\bs k}\left( a^+_{\bs k} a_{\bs k}
+ b^+_{\bs k} b_{\bs k}\right)$.
With account for this the partition function of an ideal gas $Z_0$ reads
\begin{eqnarray}
\label{eq:z0-part-func-1}
Z_0 &=&  \frac{\beta}{2\pi i}\int_0^{2\pi i} d\alpha \, e^{- \beta \alpha N_I} \,
{\rm Tr}\left[ e^{- \beta H}\,
e^{\beta \alpha \left(\hat N^{(-)} - \hat N^{(+)} \right)} \, \right]
\\
&=&  \frac{\beta}{2\pi i}\int_0^{2\pi i} d\alpha \, e^{ - \beta \alpha N_I}
\sum_{\{N^{(-)}_{\bs k}\},\{N^{(+)}_{\bs k}\}}
e^{ \beta \alpha \sum_{\bs k} \left(N^{(-)}_{\bs k} - N^{(+)}_{\bs k} \right)
 - \beta \sum_{\bs k} \omega_{\bs k} \left(N^{(-)}_{\bs k} + N^{(+)}_{\bs k} \right) }
\nonumber \\
&=&
\frac{\beta}{2\pi i}\int_0^{2\pi i} d\alpha \, e^{ - \beta \alpha N_I}
\sum_{\{N^{(-)}_k\}} \prod_k e^{ - \beta (\omega_k - \alpha) N^{(-)}_k }
\sum_{\{N^{(+)}_k\}} \prod_k e^{ - \beta (\omega_k + \alpha) N^{(+)}_k }  \,,
\label{eq:z0-part-func}
\end{eqnarray}
where the product means:
$\prod_k f(N_k) = f(N_1)\,f(N_2)\, \ldots \, f(N_{\rm max})$.
Therefore, equation (\ref{eq:z0-part-func}) leads to
\begin{eqnarray}
Z_0 =  \frac{\beta}{2\pi i}\int_0^{2\pi i} d\alpha \, e^{ - \beta \alpha N_I}
\prod_k \sum_{N^{(-)}_k = 0}^\infty e^{ - \beta (\omega_k - \alpha) N^{(-)}_k }
\prod_k \sum_{N^{(+)}_k = 0}^\infty e^{ - \beta (\omega_k + \alpha) N^{(+)}_k }  \,,
\label{eq:z0-part-func2}
\end{eqnarray}
where in the case of a bosonic system the sums can be easily calculated
\begin{eqnarray}
\sum_{N_k = 0}^\infty e^{ - \beta (\omega_k - \alpha) N_k }
\,=\, \left[ 1 - e^{ - \beta (\omega_k - \alpha) } \right]^{-1} \,
\equiv \, z(\omega_k,\alpha) \,.
\label{eq:sp-part-func}
\end{eqnarray}
Applying this result to eq.~(\ref{eq:z0-part-func2}), we obtain
\begin{eqnarray}
Z_0 \,=\,  \frac{\beta}{2\pi i}\int_0^{2\pi i} d\alpha \,
e^{ - \beta \alpha N_I + \sum_k \, \ln{ \left[ z(\omega_k,\alpha) \,
z(\omega_k,-\alpha)\right] } }  \,.
\label{eq:z0-part-func3}
\end{eqnarray}
The integral on the right of this equality can be calculated by
the steepest descent method.
We define the function
\begin{equation}
w(\alpha) \,=\, - \beta \alpha + \frac{1}{ N_I} \sum_k \,
\ln{ \left[ z(\omega_k,\alpha) \, z(\omega_k,-\alpha)\right] } \,.
\label{eq:alpha-exponent}
\end{equation}
The midpoint $\alpha_0$, for the expansion of the function $w(\alpha)$ around it,
is determined from equation $w'(\alpha) = 0$.
We get
\begin{equation}
- \beta + \frac{1}{ N_I} \sum_k \,\left[
\frac{1}{z(\omega_k,\alpha)} \frac{\partial z(\omega_k,\alpha)}{\partial \alpha}
+ \frac{1}{z(\omega_k,-\alpha)} \frac{\partial z(\omega_k,-\alpha)}{\partial \alpha}
\right]  \,=\, 0 \,.
\label{eq:alpha-exponent2}
\end{equation}
Taking into account the definition (\ref{eq:sp-part-func}) of the function
$z(\omega_k,\alpha)$, we rewrite eq.~(\ref{eq:alpha-exponent2}) as
\begin{equation}
N_I \,=\, - \frac{\partial}{\partial \alpha} \sum_k \,\left[
\ln{ \left(1 - e^{\beta(\alpha -  \omega_k)}\right)}
+ \ln{ \left(1 - e^{\beta(- \alpha - \omega_k)}\right)} \right] \,.
\label{eq:eq-for-ni}
\end{equation}
In the limit $V \to \infty$, the sum is converted into an integral, and
for a given particle-number density $n_I = N_I/V$ we get equation for $\alpha$:
\begin{equation}
n_I \,=\, \int \frac{d^3p}{(2\pi)^3} \,\left[
     \frac{1}{e^{ \beta (\omega_k - \alpha)} - 1} \,
-\, \frac{1}{e^{ \beta (\omega_k + \alpha)} - 1} \right] \,.
\label{eq:eq-for-ni-3}
\end{equation}
The solution of this equation with respect to $\alpha$ gives the function
$\alpha = \alpha_0(T,n_I)$, which we use in the next step, where we expand
the function $w(\alpha)$ defined in eq.~(\ref{eq:alpha-exponent}).
Indeed, for approximate integration in eq.~(\ref{eq:z0-part-func3}) we take
the expansion of $w(\alpha)$ around the $\alpha_0$ to the first derivative:
\begin{equation}
w(\alpha) \,\approx\, w(\alpha_0) \,.
\label{eq:alpha-exponent3}
\end{equation}
Then the approximate expression for the partition function
(\ref{eq:z0-part-func3}) has the form
\begin{eqnarray}
Z_0 & \approx & \frac{\beta}{2\pi i}\int_0^{2\pi i} d\alpha \,
e^{ - N_I w(\alpha_0)} \,
=\, \beta \,e^{ - N_I w(\alpha_0)} \,.
\label{eq:z0-part-func4}
\end{eqnarray}
Next, one can calculate the free energy (in the main asymptotic for $N_I$)
\begin{eqnarray}
F_0 = -T \ln{Z_0} \,=\, N_I \alpha_0 - T \sum_k \,
\ln{ \left[ z(\omega_k,\alpha_0) \, z(\omega_k,-\alpha_0)\right] } \,.
\label{eq:free-energy}
\end{eqnarray}
Then we use the thermodynamic properties of the free energy and calculate the
isospin chemical potential as the derivative of $F_0$:
\begin{eqnarray}
\mu_I \,=\, \frac{\partial F_0}{\partial N_I} \,=\,  \alpha_0 \,.
\label{eq:chem-pot}
\end{eqnarray}
Hence, we obtain that parameter $\alpha$ is nothing more than the isospin
chemical potential, which is the thermodynamic function that depends on
$(T,n_I)$, i.e. $\mu = \mu(T,n_I)$.
This dependence can be obtained by solving eq.~(\ref{eq:eq-for-ni-3})
with respect to $\mu_I$, substituting it for $\alpha$.
This equation can be written explicitly as
 \begin{eqnarray}
 n_I \,=\, \int \frac{d^3p}{(2\pi)^3} \,
\left[ f_{_{\rm BE}}(\omega_k, \mu_I) \,-\, f_{_{\rm BE}}(\omega_k, -\mu_I) \right] \,,
\label{eq:eq-for-ni-4}
\end{eqnarray}
where we define the Bose-Einstein distribution function
%
\begin{eqnarray}
f_{_{\rm BE}}(\omega, \mu) \,=\, \left[ e^{  \beta (\omega - \mu) } - 1 \right]^{-1} \,.
\label{eq:be-distr-func}
\end{eqnarray}
Therefore, in further calculations, it should be used solution of
eq.~(\ref{eq:eq-for-ni-4}), i.e. from now on the chemical potential is not
a free variable, but is a function of conserved charge and temperature,
$\mu_I = \mu_I(T,n_I)$.

Inserting the chemical potential into eq.~(\ref{eq:free-energy}) we rewrite the
free energy of the ideal gas in the following form
\begin{eqnarray}
F_0 = \mu_I N_I  - T \sum_k \,\left[ \ln{ z(\omega_k,\mu_I) }\,
+ \ln{ z(\omega_k,-\mu_I) } \right] \,.
\label{eq:free-energy2}
\end{eqnarray}
For continuous momentum ($V \to \infty$) one can write the free energy density
$\Phi_0 = F_0/V$ (compare with eq.~(\ref{eq:charge-legendre-transform-id})) as
\begin{eqnarray}
\Phi_0 =  n_I \mu_I(T,n_I) \!
+\! T \! \! \int \! \frac{d^3k}{(2\pi)^3} \! \left\{
\ln{ \! \left[ 1- e^{- \big(\omega_k - \mu_I(T,n_I) \big)/T } \right] }
\! + \ln{ \! \left[ 1- e^{- \big(\omega_k + \mu_I(T,n_I) \big)/T } \right] } \right\}\!.
\end{eqnarray}
From here we get the pressure ($\Phi = \mu_I n_I - p$)
\begin{eqnarray}
p = - T \int \frac{d^3p}{(2\pi)^3} \,\left\{
\ln{ \left[ 1 - e^{ - \beta \left(\omega_k - \mu_I(T,n_I)\right) } \right] }\,
+ \ln{ \left[ 1 - e^{ - \beta \left(\omega_k + \mu_I(T,n_I)\right) } \right] }
\right\} \,.
\label{eq:pressure}
\end{eqnarray}
%

\medskip

\noindent
{\it Particles are in a condensate phase ($T < T_{\rm c}$).}
\\
In this case, particles $N^{(-)}$ and antiparticles $N^{(+)}$ have two
fractions, thermal and condensed:
$N^{(-)} = N^{(-)}_{\rm cond} + N^{(-)}_{\rm th}$,
$N^{(+)} = N^{(+)}_{\rm cond} + N^{(+)}_{\rm th}$.
At the same time, the isospin number can also be divided into two parts,
thermal and condensed: $N_{I} = N_{I\rm cond} + N_{I\rm th}$, where
$N_{I\rm cond} = N^{(-)}_{\rm cond} - N^{(+)}_{\rm cond}$ and
$N_{I\rm th} = N^{(-)}_{\rm th} - N^{(+)}_{\rm th}$.
Therefore, in the condensate phase $N_I$ written as:
\[ N_{I} \,=\, N_{I\rm cond} + N_{I\rm th} \,.
\]
The Hamiltonian in the condensate phase has the form
$H = \varepsilon_0 N_{\rm cond}
+ \sum_{\bs k \ne 0} \omega_k(a^+_{\bs k} a_{\bs k} + b^+_{\bs k} b_{\bs k})$.
With account for the condensate in the system, the partition function $Z$ can
be written as (for comparison, see eq.~(\ref{eq:charge-id-gce-part-func-cond-2})):
\begin{eqnarray}
Z = e^{-\beta\epsilon_0 N_{\rm cond}} \frac{\beta}{2\pi i}\int_0^{2\pi i}
d\alpha \,
e^{- \beta \alpha N_{I\rm th}} \, {\rm Tr}\left[ e^{- \beta H_{\rm th}}\,
e^{\beta \alpha \left(\hat N^{(-)}_{\rm th} - \hat N^{(+)}_{\rm th} \right)} \,
\right] \,.
\label{eq:z0-kronecker-1}
\end{eqnarray}
where $\hat N^{(-)}_{\rm th} = \sum_{\bs k \ne 0} a^+_{\bs k} a_{\bs k}$,
$\hat N^{(+)}_{\rm th} = \sum_{\bs k \ne 0} b^+_{\bs k} b_{\bs k}$ and
$N_{I\rm th}=N_I-\left(N^{(-)}_{\rm cond} - N^{(+)}_{\rm cond}\right)$.
Comparing the partition function $Z$ given in eq.~(\ref{eq:z0-kronecker-1})
with that for the thermal phase (\ref{eq:z0-part-func-1}), it is evident that
the difference belongs only to the factor
$\exp{(-\beta\epsilon_0 N_{\rm cond})} $, which is simply a C-number.
Therefore, we can perform the calculation of the ``thermal'' integral as it
was done in the previous section for $T > T_{\rm c}$.
Conducting calculations, we are coming to eq.~(\ref{eq:eq-for-ni-3}) that
determines the value of mid-alpha.
The equation reads
\begin{equation}
n_{I{\rm th}} \,=\, \int \frac{d^3p}{(2\pi)^3} \,\left[
     \frac{1}{e^{ \beta (\omega_k - \alpha)} - 1} \,
-\, \frac{1}{e^{ \beta (\omega_k + \alpha)} - 1} \right] \,.
\label{eq:eq-for-ni-3-cond}
\end{equation}
Remember, we are considering the condensate phase.
If both components are in the condensate phase, then it should be valid
at the same time:  $m - \alpha = 0$ and $m + \alpha = 0$.
This is $m = 0$ and $\alpha = 0$.
But we study the system of massive particles, so there are two opportunities:
$\alpha = m$ or $\alpha = - m$.
We choose the first version for the current calculations, $\alpha = m$.
The latter means that condensate is created only by particles, i.e.,
$N_{\rm cond} = N_{\rm cond}^{(-)} = N_{I\rm cond}$
and $N_{\rm cond}^{(+)} = 0$.
Then, eq.~(\ref{eq:eq-for-ni-3-cond}) should be rewritten as
\begin{equation}
n_{I{\rm th}} \,=\, n_{\rm lim}(T) -
\int \frac{d^3p}{(2\pi)^3} \, f_{_{\rm BE}}(\omega_k, - m)
\quad  {\rm with}  \quad
n_{\rm lim}(T) \,=\, \int \frac{d^3p}{(2\pi)^3} \, f_{_{\rm BE}}(\omega_k, m) \,,
\label{eq:eq-for-ni-3-cond-2}
\end{equation}
where $n_{\rm lim}(T)$ is the critical curve that separates condensate states
from pure thermal states in the $(T,n)$-plane.
So, the total isospin (charge) density reads
\begin{equation}
n_I \,=\, n_{\rm cond}(T) \,+\, n_{I{\rm th}}(T) \,.
\label{eq:ni}
\end{equation}
Next, taking the partition function from eq.~(\ref{eq:z0-kronecker-1}),
with account for result (\ref{eq:free-energy}) of the calculation of the
thermal free energy, we calculate the free energy ($F = - T \ln Z$):
\begin{eqnarray}
F \,=\, \epsilon_0 N_{\rm cond} + m N_{I\rm th}
- T \sum_k \, \ln{ \left[ z(\omega_k,m) \, z(\omega_k,-m)\right] } \,.
\label{eq:free-energy-cond}
\end{eqnarray}
We take into account that $\epsilon_0 = m$ and $N_{\rm cond} = N_{I\rm cond}$,
therefore $N_{\rm cond} + N_{I\rm th} = N_I$.
Then, for continuous momentum ($V \to \infty$) we can write
eq.~(\ref{eq:free-energy-cond}) as
\begin{eqnarray}
\Phi \,=\, m \, n_I \,+\, T \int \frac{d^3k}{(2\pi)^3}
\left\{\ln{ \left[ 1- e^{- \big(\omega_k - m \big)/T } \right] }
+ \ln{ \left[ 1- e^{- \big(\omega_k + m \big)/T } \right] } \right\}  \,,
\end{eqnarray}
where $\Phi = F/V$ is a density of the free energy.
We see that this expression coincides with the free energy density obtained in
eq.~(\ref{eq:cond-dens-free-energy}).


\end{document}